\documentclass[11pt,a4paper]{article}

\usepackage[T1]{fontenc}
\usepackage[utf8]{inputenc}
\usepackage[left=3.0cm,right=3.0cm,top=3.2cm,bottom=3.2cm]{geometry}
\usepackage{amsmath,amssymb,amsthm,mathrsfs}
\usepackage{bbm}
\usepackage{booktabs}
\usepackage{xcolor}
\usepackage[numbers,sort&compress]{natbib}
\usepackage{hyperref}
\usepackage{titlesec}
\usepackage{caption}
\usepackage{float}
\usepackage{pgfplots}
\pgfplotsset{compat=1.17}
\usepackage{tikz}
\usetikzlibrary{arrows.meta,positioning,calc}
\usepackage{tcolorbox}
\tcbuselibrary{skins,breakable}
\usepackage{array}

\definecolor{qblue}{RGB}{0,70,140}
\definecolor{qorange}{RGB}{210,130,0}
\definecolor{qgreen}{RGB}{0,158,115}
\definecolor{qred}{RGB}{213,94,0}
\definecolor{qpurple}{RGB}{148,103,189}
\definecolor{qgray}{RGB}{100,100,100}
\definecolor{qboxbg}{RGB}{237,244,251}

\hypersetup{colorlinks,linkcolor=qblue,citecolor=qblue,urlcolor=qblue}

\titleformat{\section}{\large\bfseries}{\thesection}{0.7em}{}
\titleformat{\subsection}{\normalsize\bfseries}{\thesubsection}{0.6em}{}
\titleformat{\subsubsection}{\normalsize\itshape}{\thesubsubsection}{0.5em}{}
\titlespacing{\section}{0pt}{10pt plus2pt}{4pt plus1pt}
\titlespacing{\subsection}{0pt}{7pt plus1pt}{2pt plus1pt}

\theoremstyle{remark}
\newtheorem{remark}{Remark}

\newtcolorbox{keyresult}[1][]{
  colback=qboxbg, colframe=qblue,
  fonttitle=\bfseries\small, title={#1},
  boxrule=0.5pt, left=6pt, right=6pt, top=4pt, bottom=4pt,
  breakable, arc=2pt}

\pgfplotsset{
  qplot/.style={
    grid=major,
    grid style={line width=0.25pt, draw=qgray!25},
    tick label style={font=\small},
    label style={font=\small},
    tick align=inside,
    line width=0.9pt,
  }
}

\newcommand{\Pket}{|{P}\rangle}
\newcommand{\Rket}{|{R}\rangle}
\newcommand{\DeltaE}{\Delta E}
\newcommand{\Jz}{\hat{J}_z}
\newcommand{\Jx}{\hat{J}_x}
\newcommand{\hP}{\hat{P}}
\newcommand{\Gcoll}{G_{\rm loc}}
\newcommand{\Geig}{G_{01}}
\newcommand{\mstar}{m_*}
\newcommand{\tr}{\operatorname{tr}}
\newcommand{\Jol}{J_{01}}

\begin{document}

\begin{center}
\vspace*{8pt}
{\LARGE\bfseries
Which Coherence Decoheres?\\[4pt]
Basis-Dependent Decoherence Rates\\[4pt]
in Symmetry-Broken Collective Spin Systems}\\[16pt]
{\normalsize\bfseries Stavros Mouslopoulos}\\[4pt]
{\small\itshape
Faculty of Science and Engineering,
University of Nottingham Ningbo China\\
Ningbo 315100, China}\\[3pt]
{\small\texttt{Stavros.Mouslopoulos@nottingham.edu.cn}}\\[12pt]
\rule{0.8\linewidth}{0.4pt}
\end{center}

\begin{abstract}
\noindent
For a collective spin system in a symmetry-broken phase, there exist
two natural bases: the localised states $\{|P\rangle,|R\rangle\}$
that maximise the order parameter within the low-energy doublet, and the energy eigenstates
$\{|E_0\rangle,|E_1\rangle\}$ that diagonalise the Hamiltonian.
Both bases yield well-controlled Lindblad dephasing rates for their
respective coherences, and in the mesoscopic quantum regime, those
rates differ.
The exact diagonal Redfield coefficient for $\rho_{PR}$ is
$\gamma_\phi(G_{01} + J_{01}^2)$; the mean-field (spin-coherent)
approximation gives the larger rate $\gamma_\phi\Gcoll$ where $\Gcoll=(N\mstar)^2/2$,
which governs the pure exponential decay of $\mathrm{Re}(\rho_{PR})$.
The energy-eigenstate coherence $\rho_{01}$ decays at rate
$\gamma_\phi\Geig$ where
$\Geig=\tfrac{1}{2}(\langle E_0|\Jz^2|E_0\rangle
                   +\langle E_1|\Jz^2|E_1\rangle)$.
These two questions yield different answers in two distinct senses:
the \emph{geometric} ratio $\eta_{\rm MF}=\Gcoll/\Geig$ approaches exactly
$2$ as $N\rightarrow\infty$ (a model-independent statement about
matrix elements, valid even where the secular approximation fails);
and at \emph{finite} $N$ in the LMG model, $\eta_{\rm MF}$ rises to a peak
$\eta_{\rm MF}\approx2.42$ within the mesoscopic secular window
($2\DeltaE\cdot T_2\gg1$, setting $\hbar=1$), where both rates are simultaneously
well-defined as exponential decay constants.
The discrepancy in physical decay rates is therefore strictly a
mesoscopic finite-$N$ effect; the universal $\eta_{\rm MF}=2$ is a
geometric statement that survives the breakdown of the secular
approximation in the thermodynamic limit.
In the thermodynamic limit ($N\rightarrow\infty$), the secular
approximation fails, the doublet becomes degenerate, and the
\emph{decaying components} of both coherences converge exactly to
the classical macroscopic (pointer-basis) rate $\gamma_\phi\Gcoll$,
while
$\mathrm{Re}(\rho_{01})$ approaches a quasi-steady state (metastable plateau) within the doublet.
At finite $N$, however, the secular approximation remains valid,
and the two coherences decay at genuinely different rates.
For quantum technologies --- spin squeezing, quantum Fisher
information, Leggett-Garg tests --- this provides a quantitative
quantum advantage \emph{conditional on the protocol being sensitive
to the eigenstate coherence $\rho_{01}$ and remaining within the
ground-state doublet}.
Two distinct protection factors are relevant:
$\eta_{\rm MF}=\Gcoll/\Geig\approx2.42$ (peak), which quantifies the
advantage over the classical mean-field pointer-state estimate
$\gamma_\phi\Gcoll$;
and $\eta_{\rm exact}=(G_{01}+J_{01}^2)/G_{01}\approx1.86$,
which is the true basis-dependent physical protection factor ---
the exact ratio of the pointer-state decay rate $\gamma_\phi(G_{01}+J_{01}^2)$
to the eigenstate rate $\gamma_\phi G_{01}$.
We demonstrate this protected three-regime structure in the
Lipkin-Meshkov-Glick model via exact diagonalisation and provide the
precise algebraic origin of the discrepancy via the $\mathbb{Z}_2$
parity of the Lindblad operator.
\end{abstract}

\vspace{4pt}
\noindent\rule{\linewidth}{0.3pt}
\vspace{4pt}

\section{Introduction}
\label{sec:intro}

The decoherence rate of a collective spin system is not a single
number. It is a basis-dependent statement about which coherence is
decaying, and at what rate.

This observation has a concrete consequence. In the ordered phase
of any $\mathbb{Z}_2$-symmetric collective spin model, two
Lindblad-based calculations yield rates that differ by a factor
approaching $2$ as $N\rightarrow\infty$ and reaching $2.35$--$2.42$
near the quantum-critical crossover at finite $N$.
Neither calculation is wrong. They answer different questions.
The localised-state calculation correctly describes the dephasing
of $\rho_{PR}=\langle P|\rho|R\rangle$, the coherence between
the exact symmetry-broken pointer states. The eigenstate calculation correctly
describes the dephasing of $\rho_{01}=\langle E_0|\rho|E_1\rangle$,
the coherence between energy eigenstates.

These are not the same object (see Eq.~\eqref{eq:rho01_relation}
for their relation);
they decay at \emph{different} rates in the mesoscopic secular
window (e.g., $N\approx250$--$430$) where the secular approximation
remains robust ($2\DeltaE\cdot T_2\gg1$) and both rates are
simultaneously well-defined.
At finite $N$ the discrepancy is amplified by near-critical
wavefunction delocalisation.

This discrepancy is strictly a mesoscopic quantum phenomenon. In the
macroscopic classical limit ($N\rightarrow\infty$), the secular
approximation breaks down, the ground doublet becomes degenerate, and
the exact Lindblad dynamics force the decaying components in
\emph{both} bases to converge to the fast rate $\gamma_\phi\Gcoll$,
while $\mathrm{Re}(\rho_{01})$ approaches a quasi-steady state (metastable plateau) within the doublet.
The two pictures are not in contradiction; rather, the convergence of
the physical rates marks the emergence of classical order-parameter
relaxation (Section~\ref{sec:limit}).

The rate difference exists solely in the finite-$N$ regime, where
quantum tunnelling ($\DeltaE>0$) dynamically protects the eigenstate
coherence. For protocols sensitive to the eigenstate coherence
$\rho_{01}$ and operating effectively within the ground-state
doublet, the relevant decoherence rate is therefore
$\gamma_\phi\Geig$ rather than the localised-state estimate
$\gamma_\phi\Gcoll$, with quantum-coherence lifetimes longer by
a factor of $\eta_{\rm exact}\approx1.86$ at $N=370$ (the true
physical basis-dependent protection), or up to $\eta_{\rm MF}\approx2.42$
when comparing against the classical mean-field pointer-state estimate.
Note that this dynamically protected secular window
($N\approx250$--$430$) is broader than, and fully encompasses, the
thermal ``Goldilocks'' zone ($N\approx250$--$370$) defined in the
companion paper~\cite{mouslopoulos2026}, where the additional
constraint $\DeltaE \gtrsim k_B T$ is required to suppress thermal
noise in experimental implementations.\footnote{The temperature is
set by $\DeltaE(N=370)=k_BT$, giving $T\approx10$~nK for
BEC-relevant parameters~\cite{mouslopoulos2026}. The thermal
Goldilocks zone $\DeltaE\gtrsim k_BT$ corresponds to $N\lesssim370$;
combining this with the secular condition $2\DeltaE\cdot T_2\gg1$
(satisfied for $N\gtrsim250$) defines the window $N\approx250$--$370$
where both thermal stability and quantum coherence are simultaneously
optimised.}

\textit{Relation to prior work.} The basis dependence of decoherence
is rooted in Zurek's einselection programme~\cite{Zurek2003}, which
identifies the pointer basis as the set of states stable under
system-bath coupling.
The Parity Proposition of Section~\ref{sec:theorem} is conceptually
adjacent to the decoherence-free-subspace
construction~\cite{Lidar1998}, where a discrete symmetry of the
system-bath coupling protects coherences within a symmetry sector;
the present mechanism differs in that the eigenstate coherence
$\rho_{01}$ is not strictly protected but rather decays at a slower
rate than the localised-state coherence by the universal factor
$\eta_{\rm MF}\rightarrow2$.
Dissipative dynamics in the LMG and Dicke models, including pointer-state
selection by collective operators, symmetry-protected decoherence,
and decoherence signatures of excited-state quantum phase transitions,
have been studied extensively~\cite{Morrison2008,Trenkwalder2016,Botet1982,Relano2008}.
The factor-of-$\eta_{\rm MF}$ distinction analysed here is, to our knowledge,
not previously isolated as a basis-dependent statement about two
\emph{simultaneously well-defined} dephasing rates in the
mesoscopic secular window: prior work has typically focused on the
order-parameter rate $\gamma_\phi\Gcoll$ for symmetry-broken
phases~\cite{Tegmark2000} or on the eigenstate rate $\gamma_\phi\Geig$
for spectroscopic protocols~\cite{Wineland1992,Pezze2018} but has
not unified the two within a single framework with the explicit
parity-based proposition we provide in Section~\ref{sec:theorem}.

The physical question is which coherence governs the observable of
interest. For classical order-parameter relaxation, the localised
basis is natural. For quantum-information protocols --- spin
squeezing~\cite{Wineland1992,Kitagawa1993}, quantum Fisher
information~\cite{Pezze2018}, Leggett-Garg inequality
tests~\cite{Leggett1985,Emary2014} --- the relevant coherence is
$\rho_{01}$, and the slower eigenstate rate is correct.
Two distinct protection factors are relevant in this paper:
\begin{align}
  \eta_{\rm MF}(N,\Gamma/J)
  &\equiv \frac{\Gcoll}{\Geig}
  = \frac{(N\mstar)^2/2}{\tfrac{1}{2}
    \bigl(\langle E_0|\Jz^2|E_0\rangle
         +\langle E_1|\Jz^2|E_1\rangle\bigr)},
  \label{eq:eta_def}\\
  \eta_{\rm exact}(N,\Gamma/J)
  &\equiv \frac{\Geig + J_{01}^2}{\Geig}
  = 1 + \frac{J_{01}^2}{\Geig}.
  \label{eq:eta_exact_def}
\end{align}
$\eta_{\rm MF}$ is the ratio of the classical mean-field pointer-state
rate $\gamma_\phi\Gcoll$ to the exact eigenstate rate $\gamma_\phi\Geig$;
it approaches $2$ as $N\rightarrow\infty$ and peaks at $2.42$
near the crossover of the LMG model.
$\eta_{\rm exact}$ is the ratio of the \emph{exact} pointer-state
decay rate $\gamma_\phi(\Geig+J_{01}^2)$ to the eigenstate rate;
at $N=370$ it equals $\approx1.86$.
$\eta_{\rm MF}$ decomposes exactly as $\eta_{\rm MF}=\eta_{\rm geom}\times\eta_{\rm quantum}$
(Section~\ref{sec:unified}), where $\eta_{\rm geom}=2$ is a universal
geometric factor from parity and $\eta_{\rm quantum}>1$ is a
many-body Bogoliubov enhancement reaching $1.20$ at the peak.
The thermodynamic \emph{geometric} limit $\eta_{\rm MF}\rightarrow2$ is
universal for any collective spin system satisfying two conditions:
(i)~an isolated ground-state doublet arising from spontaneous
$\mathbb{Z}_2$ symmetry breaking (so that the eigenstate expansion
Eqs.~\eqref{eq:E0}--\eqref{eq:E1} holds);
(ii)~a parity-odd Lindblad operator
($\hat{P}\hat{L}\hat{P}^\dagger=-\hat{L}$, as in
Section~\ref{sec:theorem}).
These two conditions are sufficient for the \emph{geometric} ratio
$\Gcoll/\Geig\rightarrow2$; the third condition --- the secular
approximation $2\DeltaE\cdot T_2\gg1$ --- is required for
$\gamma_\phi\Geig$ to be a physically meaningful exponential rate,
but is not needed for the geometric statement.
The specific peak value $\eta_{\rm MF}\approx2.42$ is LMG-specific,
determined by the instanton action $S_{\rm inst}$ and the
wavefunction delocalisation near the quantum-critical crossover.

The paper is organised as follows.
Section~\ref{sec:setup} defines the two bases and coherences.
Section~\ref{sec:rates} derives both dephasing rates, including
the thermodynamic limit (Section~\ref{sec:limit}) and the
degenerate-doublet Lindblad treatment with the three-regime structure
(Section~\ref{sec:degen}).
Section~\ref{sec:finite} addresses finite-$N$ amplification.
Section~\ref{sec:lmg} presents the LMG model numerics.
Section~\ref{sec:theorem} gives the parity proposition.
Section~\ref{sec:implications} discusses implications.

\begin{table}[H]
\centering
\caption{\textbf{Glossary of Key Notation.} Summary of the primary physical parameters, geometric factors, and ratios used throughout this work.}
\label{tab:notation}
\renewcommand{\arraystretch}{1.3}
\small
\begin{tabular}{>{\raggedleft\arraybackslash}p{0.15\linewidth} p{0.75\linewidth}}
\toprule
\textbf{Symbol} & \textbf{Definition} \\
\midrule
$N, \Gamma, J$ & Number of spins, transverse field, and ferromagnetic coupling \\
$\mstar$ & Mean-field order parameter, $\sqrt{1-\Gamma^2/J^2}$ \\
$\DeltaE$ & Quantum tunnel splitting between $|E_0\rangle$ and $|E_1\rangle$ \\
$\gamma_\phi$ & Markovian collective dephasing rate \\
$\Gcoll$ & Mean-field geometric factor, $(N\mstar)^2/2$ \\
$\Geig$ & Exact eigenstate geometric factor, $\frac{1}{2}(\langle E_0|\Jz^2|E_0\rangle + \langle E_1|\Jz^2|E_1\rangle)$ \\
$\Jol$ & Exact off-diagonal matrix element, $\langle E_0|\Jz|E_1\rangle$ \\
$\eta_{\rm MF}$ & Advantage over classical mean-field estimate, $\Gcoll/\Geig$ \\
$\eta_{\rm exact}$ & Exact basis-dependent physical protection factor, $(\Geig+\Jol^2)/\Geig$ \\
\bottomrule
\end{tabular}
\end{table}

\section{Two Bases, Two Coherences}
\label{sec:setup}

Consider a collective spin-$\tfrac{1}{2}$ system of $N$ particles
with all-to-all ferromagnetic coupling $J>0$ and transverse field
$\Gamma$, described by the Kac-normalised Hamiltonian
(the $1/N$ prefactor on the coupling ensures extensive energy
in the thermodynamic limit~\cite{Kac1963})
\begin{equation}
  \hat{H} = -\frac{2J}{N}\Jz^2 - 2\Gamma\Jx,
  \label{eq:H}
\end{equation}
where $\Jz=\frac{1}{2}\sum_i\hat\sigma_i^z$ and
$\Jx=\frac{1}{2}\sum_i\hat\sigma_i^x$.
In the ordered phase ($\Gamma<J$) the $\mathbb{Z}_2$ symmetry
generated by $\hP=e^{i\pi\Jx}$ is spontaneously broken: the
mean-field ground state bifurcates into two degenerate localised
instanton vacua $|\tilde{P}\rangle$ and $|\tilde{R}\rangle$ with
$\langle \tilde{P}|\Jz|\tilde{P}\rangle=+N\mstar/2$ and
$\langle \tilde{R}|\Jz|\tilde{R}\rangle=-N\mstar/2$,
where $\mstar=\sqrt{1-\Gamma^2/J^2}$ is the order parameter.

For finite $N$, the true eigenstates of $\hat{H}$ are\footnote{Readers comparing this work with the companion paper~\cite{mouslopoulos2026} should note a necessary shift in state notation. In Ref.~\cite{mouslopoulos2026}, $|P\rangle$ and $|R\rangle$ denote the non-orthogonal mean-field spin-coherent states. Because the central thesis of the present work relies on the precise algebraic distinction between exact finite-$N$ states and mean-field approximations, we strictly reserve $|P\rangle$ and $|R\rangle$ for the exact orthogonal pointer states defined in the energy doublet, using $|\tilde{P}\rangle$ and $|\tilde{R}\rangle$ (or $|P_{\rm SCS}\rangle$ and $|R_{\rm SCS}\rangle$) for the classical spin-coherent states.}:
\begin{align}
  |E_0\rangle &= \frac{|\tilde{P}\rangle+|\tilde{R}\rangle}{\sqrt{2(1+S)}}
               = \frac{|\tilde{P}\rangle+|\tilde{R}\rangle}{\sqrt{2}}
               + \mathcal{O}(S),
  \label{eq:E0}\\
  |E_1\rangle &= \frac{|\tilde{P}\rangle-|\tilde{R}\rangle}{\sqrt{2(1-S)}}
               = \frac{|\tilde{P}\rangle-|\tilde{R}\rangle}{\sqrt{2}}
               + \mathcal{O}(S),
  \label{eq:E1}
\end{align}
where $S_{\rm inst} = \text{arctanh}(m_*)-m_*$ is the WKB instanton
action~\cite{mouslopoulos2026} governing the exponentially small tunnel
splitting $\DeltaE=E_1-E_0 = C_0 N^{1/2}e^{-NS_{\rm inst}}$
(at $\Gamma/J=0.95$: $S_{\rm inst}=0.010787$, giving $\DeltaE\approx1310$~rad/s
at $N=370$ from exact diagonalisation);
and $S\equiv\langle\tilde{P}|\tilde{R}\rangle = (\Gamma/J)^N \approx 5.7\times10^{-9}$
is the instanton-vacuum overlap at the benchmark.
These two small quantities are physically distinct:
$e^{-NS_{\rm inst}}\approx0.018$ governs the energy splitting $\DeltaE$,
while $S\approx5.7\times10^{-9}$ governs the $\mathcal{O}(S)$
wavefunction correction in Eqs.~\eqref{eq:E0}--\eqref{eq:E1};
they differ by a factor $\approx3\times10^6$.
Their parity eigenvalues are $p_0=+1$ (even) and $p_1=-1$ (odd).

We define the exact orthogonal pointer states of the low-energy doublet as $|P\rangle = (|E_0\rangle+|E_1\rangle)/\sqrt{2}$ and $|R\rangle = (|E_0\rangle-|E_1\rangle)/\sqrt{2}$.

The system supports two distinct coherences:
\begin{align}
  \rho_{PR} &\equiv \langle P|\rho|R\rangle
  \quad\text{(localised-state coherence)},
  \label{eq:rhoPR}\\
  \rho_{01} &\equiv \langle E_0|\rho|E_1\rangle
  \quad\text{(eigenstate coherence)}.
  \label{eq:rho01}
\end{align}
They are generically distinct.
Using the exact orthogonal pointer states, one finds the exact algebraic identity:
\begin{equation}
  \rho_{01} = \tfrac{1}{2}(\rho_{PP}-\rho_{RR})
             -\tfrac{1}{2}(\rho_{PR}-\rho_{RP}).
  \label{eq:rho01_relation}
\end{equation}
Under the additional conditions $\rho_{PP}=\rho_{RR}$ (equal
localised populations) and $\rho_{PR}$ purely imaginary
($\rho_{PR}=iy$, $\rho_{RP}=-iy$), one finds $\rho_{01}=-\rho_{PR}$:
equal in magnitude but opposite in sign, differing by an overall
relative phase of $\pi$ that leaves the decay rates invariant
but profoundly alters the physical observables of the coherence.
All three conditions (well-developed double-well, equal populations,
purely imaginary $\rho_{PR}$) are special cases; the coherences are
generically distinct.

\begin{table}[H]
\centering
\caption{\textbf{Taxonomy of Quantum and Classical States.} Summary of the five distinct state types discussed in this work, classified by their physical regime (Quantum, Semi-Classical, Classical). The instanton vacua $|\tilde{P}\rangle$, $|\tilde{R}\rangle$ are the WKB saddle-point states that mediate between the exact quantum pointer states and the classical mean-field product states; their overlap $S=\langle\tilde{P}|\tilde{R}\rangle=(\Gamma/J)^N$ governs the $\mathcal{O}(S)$ corrections in the eigenstate expansions. The distinction between the exact quantum pointer states and the classical mean-field SCS is the physical origin of the $\eta_{\rm exact}$ vs $\eta_{\rm MF}$ discrepancy.}
\label{tab:state_taxonomy}
\renewcommand{\arraystretch}{1.4}
\small
\begin{tabular}{>{\raggedright\bfseries}p{0.18\linewidth} >{\raggedright}p{0.25\linewidth} >{\raggedright}p{0.25\linewidth} >{\raggedright\arraybackslash}p{0.22\linewidth}}
\toprule
State / Basis & Definition \& Relation & Mesoscopic Regime \newline ($N \approx 250$--$430$) & Thermodynamic Limit \newline ($N \rightarrow \infty$) \\
\midrule

Energy Eigenstates \newline \textcolor{qblue}{\scriptsize [EXACT QUANTUM]} \newline $\{|E_0\rangle, |E_1\rangle\}$ &
Exact eigenstates of $\hat{H}$. \newline
Definite $\mathbb{Z}_2$ parity ($p_0=+1$, $p_1=-1$). \newline
$|E_{0,1}\rangle = \frac{1}{\sqrt{2}}(|P\rangle \pm |R\rangle)$. &
Form a quasi-degenerate doublet ($\Delta E > 0$). \newline
\textcolor{qblue}{Decay rate: $\gamma_\phi G_{01}$.} &
Exactly degenerate ($\Delta E \to 0$). Secular approx.\ fails. \\

\midrule

Exact Pointer States \newline \textcolor{qblue}{\scriptsize [EXACT QUANTUM]} \newline $\{|P\rangle, |R\rangle\}$ &
Bath-selected states; exact eigenstates of $\hat{J}_z$ \emph{within the doublet}. \newline
$|P,R\rangle = \frac{1}{\sqrt{2}}(|E_0\rangle \pm |E_1\rangle)$. \newline
$|P\rangle = |\tilde{P}\rangle + \mathcal{O}(S)$. &
Macroscopically distinct but spin-squeezed by Bogoliubov depletion: $\langle\hat{J}_z\rangle = J_{01} < N m_*/2$. \newline
\textcolor{qblue}{Redfield coefficient: $\gamma_\phi(G_{01} + J_{01}^2)$.} &
Converge to instanton vacua ($\mathcal{O}(S)\to0$) and thence to classical mean-field states. \\

\midrule

Instanton Vacua \newline (WKB) \newline \textcolor{qpurple}{\scriptsize[SEMI-CLASSICAL]} \newline $\{|\tilde{P}\rangle, |\tilde{R}\rangle\}$ &
WKB saddle-point states; non-orthogonal ($\langle\tilde{P}|\tilde{R}\rangle = S = (\Gamma/J)^N$). \newline
$\langle\tilde{P}|\hat{J}_z|\tilde{P}\rangle = +Nm_*/2$ exactly (\textbf{This is the definition of the WKB saddle-point state's center.}). \newline
Appear in eigenstate expansions: $|E_{0,1}\rangle = \frac{|\tilde{P}\rangle \pm |\tilde{R}\rangle}{\sqrt{2}} + \mathcal{O}(S)$. &
Approximate exact pointer states with correction $\mathcal{O}(S) \approx 5.7\times10^{-9}$. \newline
Differ from SCS states by Bogoliubov depletion $\Delta_{\rm zp}\sim\mathcal{O}(1)$. \newline
Not eigenstates of $\hat{H}$. &
$S\to0$: become orthogonal and converge to exact pointer states. Still differ from SCS by $\mathcal{O}(N^{-1})$. \\

\midrule

Mean-Field States \newline (Spin-Coherent) \newline \textcolor{qred}{\scriptsize [CLASSICAL]} \newline $\{|P_{\rm SCS}\rangle, |R_{\rm SCS}\rangle\}$ &
Classical product states minimising the mean-field free energy. \newline
$\langle\hat{J}_z\rangle = \pm N m_*/2$ \textbf{exactly by construction}. \newline
Differ from $|\tilde{P}\rangle$ by Bogoliubov corrections $\mathcal{O}(N^{-1})$. &
Overestimate true pointer magnetisation by $\Delta_{\rm zp}\sim\mathcal{O}(1)$, inflating the decay rate by $\approx 26\%$. \newline
\textcolor{qblue}{Decay rate: $\gamma_\phi G_{\rm loc}$.} &
Become the exact ground states of the system; corrections vanish. \\

\midrule

Dicke States \newline \textcolor{qgray}{\scriptsize [MICROSCOPIC BASIS]} \newline $\{|J_t, m\rangle\}$ &
Exact eigenstates of $\hat{J}^2$ and $\hat{J}_z$ with eigenvalues $J_t(J_t+1)$ and $m$ ($J_t=N/2$). \newline
Computational basis in which $\hat{H}$ is tridiagonal. &
Pointer states are localised wavepackets spanning many $m$-states around $\pm J_{01}$. &
Form a continuous macroscopic variable $m_z = m/(N/2)$. \\

\bottomrule
\end{tabular}
\end{table}
\section{Two Dephasing Rates}
\label{sec:rates}

Subject the system to Markovian collective dephasing with Lindblad
operator $\hat{L}=\sqrt{\gamma_\phi}\Jz$:
\begin{equation}
  \dot\rho = -i[\hat{H},\rho]
  +\gamma_\phi\!\left(\Jz\rho\Jz
  - \tfrac{1}{2}\Jz^2\rho - \tfrac{1}{2}\rho\Jz^2\right).
  \label{eq:lindblad}
\end{equation}
We define two dimensionless geometric factors:
\begin{align}
  \Gcoll &\equiv \frac{(N\mstar)^2}{2},
  \label{eq:Gloc_def}\\
  \Geig  &\equiv \frac{1}{2}\!\left(
    \langle E_0|\Jz^2|E_0\rangle
   +\langle E_1|\Jz^2|E_1\rangle\right).
  \label{eq:Geig_def}
\end{align}
The mean-field dephasing rate is $\gamma_\phi\Gcoll$; the exact eigenstate rate is $\gamma_\phi\Geig$.

\subsection{The unified dephasing rate formula}
\label{sec:unified}

For any Hermitian Lindblad operator $\hat{L}=\sqrt{\gamma_\phi}\Jz$,
the dephasing rate of the coherence $\rho_{AB}=\langle A|\rho|B\rangle$
between any two states $|A\rangle$ and $|B\rangle$ follows directly
from projecting Eq.~\eqref{eq:lindblad}:
\begin{equation}
  \Gamma_{AB}
  = \frac{\gamma_\phi}{2}\Bigl[
      \langle A|\Jz^2|A\rangle
     +\langle B|\Jz^2|B\rangle
     -2\langle A|\Jz|A\rangle\langle B|\Jz|B\rangle
    \Bigr].
  \label{eq:Gamma_AB}
\end{equation}

Equation~\eqref{eq:Gamma_AB} is the \emph{exact diagonal Redfield
coefficient} --- the coefficient of $\rho_{AB}$ in the full
projected equation $\dot\rho_{AB} = -i\omega_{AB}\rho_{AB}
-\Gamma_{AB}\rho_{AB} + (\text{off-diagonal couplings})$.
It requires no approximation to compute; in particular, the
$\langle A|\Jz^2|A\rangle$ terms implicitly include all leakage
to higher states via the completeness identity
$\langle A|\Jz^2|A\rangle = \sum_k|\langle A|\Jz|k\rangle|^2$
(Section~\ref{sec:degen}, Eq.~\eqref{eq:Jz2_exact}).
However, obtaining a \emph{pure exponential decay} $\dot\rho_{AB}=-\Gamma_{AB}\rho_{AB}$
requires additional approximations to eliminate these off-diagonal
couplings~\cite{BreuerPetruccione2002}.
For the eigenstate coherence $\rho_{01}$, the secular approximation drops the
$\gamma_\phi|J_{01}|^2$ coupling to $\rho_{10}$ (Corollary~2,
Section~\ref{sec:theorem}), while population-coherence decoupling is
exact by parity (Corollary~1).
For the localised coherence $\rho_{PR}$, a mean-field (SCS) approximation is used, which effectively closes the dynamics to the two-state pointer sector. Under this approximation, the two-channel structure detailed in Appendix~\ref{app:rates} then reveals that only $\mathrm{Re}(\rho_{PR})$ undergoes pure exponential decay at rate $\Gamma_{PR}$, while $\mathrm{Im}(\rho_{PR})$ exhibits damped oscillations.
Equation~\eqref{eq:Gamma_AB} is nevertheless the
natural comparison formula: applied to both bases it reveals
the geometric origin of the rate difference without requiring either
approximation to be invoked, because the cross-term difference
between the two bases is an exact algebraic fact.

Applying it to the two natural bases makes the origin of the rate
difference immediately transparent:

\medskip
\noindent\textbf{Eigenstate basis $\{|E_0\rangle,|E_1\rangle\}$:}
By the Parity Proposition (Section~\ref{sec:theorem}),
$\langle E_i|\Jz|E_i\rangle=0$ exactly.
The cross-term vanishes:
\begin{equation}
  \Gamma_{01}
  = \frac{\gamma_\phi}{2}\bigl(
      \langle E_0|\Jz^2|E_0\rangle
     +\langle E_1|\Jz^2|E_1\rangle
     -2\cdot0\cdot0\bigr)
  = \gamma_\phi\Geig.
  \label{eq:rate_01_exact}
\end{equation}

\noindent\textbf{Localised basis $\{|P\rangle,|R\rangle\}$:}
Using $|P,R\rangle = (|E_0\rangle \pm |E_1\rangle)/\sqrt{2}$ and the parity proposition, the pointer states have $\langle P|\Jz|P\rangle = \frac{1}{2}(J_{01} + J_{01}^*) = \mathrm{Re}(J_{01})$. Because the LMG Hamiltonian is a real symmetric matrix, its eigenstates are purely real, making $J_{01}$ strictly real. Thus, $\langle P|\Jz|P\rangle = +J_{01}$ and similarly $\langle R|\Jz|R\rangle = -J_{01}$.
The cross-term is $-2(+J_{01})(-J_{01})=+2J_{01}^2>0$:
\begin{equation}
  \Gamma_{PR}
  = \frac{\gamma_\phi}{2}\bigl(
      \langle P|\Jz^2|P\rangle
     +\langle R|\Jz^2|R\rangle
     +2J_{01}^2\bigr)
  = \gamma_\phi\bigl(\langle P|\Jz^2|P\rangle + J_{01}^2\bigr).
  \label{eq:rate_PR_exact}
\end{equation}
\begin{keyresult}[Physical origin of the rate difference]
The entire difference between $\Gamma_{PR}$ and $\Gamma_{01}$ arises
from the cross-term $-2\langle A|\Jz|A\rangle\langle B|\Jz|B\rangle$
in Eq.~\eqref{eq:Gamma_AB}:
\begin{itemize}
  \item In the \emph{eigenstate} basis the cross-term is \emph{exactly
    zero} by parity: $\langle E_i|\Jz|E_i\rangle=0$.
  \item In the \emph{localised} basis the cross-term is
    $+2J_{01}^2\approx 2\Geig$ (large and positive), roughly
    doubling the rate.
\end{itemize}
Since $\langle P|\Jz^2|P\rangle = \frac{1}{2}(\langle E_0|\Jz^2|E_0\rangle + \langle E_1|\Jz^2|E_1\rangle) = \Geig$ (the cross-terms $\langle E_0|\Jz^2|E_1\rangle$ vanish exactly because $\Jz^2$ is parity-even), Eq.~\eqref{eq:rate_PR_exact} gives
$\Gamma_{PR} = \gamma_\phi(\Geig + \Jol^2) \approx 2\gamma_\phi\Geig$
in the thermodynamic limit where $\Jol^2\rightarrow\Geig$.
\end{keyresult}

\noindent
The remaining subsections derive the two rates from the full
Lindblad equation and state the approximations involved.

Projecting Eq.~\eqref{eq:lindblad} onto $\langle P|\cdot|R\rangle$,
the dominant dissipative contribution is:
\begin{equation}
  \dot\rho_{PR}\big|_{\rm diss}
  \simeq -\frac{\gamma_\phi}{2}
    \bigl(\langle P|\Jz|P\rangle - \langle R|\Jz|R\rangle\bigr)^2
    \rho_{PR}
  = -\gamma_\phi\Gcoll\,\rho_{PR}.
  \label{eq:rate_loc}
\end{equation}
The $\simeq$ reflects two distinct approximations, each introducing
an $\mathcal{O}(N^{-1})$ relative error in the localised rate.

\textit{Approximation 1 --- SCS mean:}
$|P\rangle$ is identified with the spin-coherent state
$|P_{\rm SCS}\rangle$, replacing the exact finite-$N$ mean
$J_{01}=\langle P|\Jz|P\rangle$ with the thermodynamic asymptote
$N\mstar/2$.
The exact pointer state $|P\rangle = (|E_0\rangle+|E_1\rangle)/\sqrt{2}$
is \emph{not} a spin-coherent state: it is spin-squeezed by
intra-well zero-point quantum fluctuations (Bogoliubov spin-wave
depletion).
As a result, $J_{01} < N\mstar/2$ with a shift of $\mathcal{O}(1)$
(numerically: $N\mstar/2 - J_{01}\approx6$--$8$ for
$N=500$--$2000$), giving a relative error of
$\mathcal{O}(1)/\mathcal{O}(N)=\mathcal{O}(N^{-1})$.
This error is completely separate from, and much larger than, the
$\mathcal{O}(S)$ overlap corrections in
Eqs.~\eqref{eq:E0}--\eqref{eq:E1}, which describe the relation
between energy eigenstates and the symmetry-broken pointer states
$|P\rangle$, $|R\rangle$ --- not between $|P\rangle$ and
$|P_{\rm SCS}\rangle$.

\textit{Approximation 2 --- SCS variance:}
The spin-coherent-state second moment is approximated as:
\begin{equation*}
  \langle P_{\rm SCS}|\Jz^2|P_{\rm SCS}\rangle
  = \left(\frac{N\mstar}{2}\right)^2 + \frac{N(1-\mstar^2)}{4}
  \simeq \left(\frac{N\mstar}{2}\right)^2,
\end{equation*}
dropping the $\mathcal{O}(N)$ spin-coherent-state variance
$N(1-\mstar^2)/4$, a relative error of
$(1-\mstar^2)/(N\mstar^2)=\mathcal{O}(N^{-1})$.

While both relative errors scale as $\mathcal{O}(N^{-1})$ and vanish in the thermodynamic limit, at the finite-$N$ benchmark ($N=370$, $\Gamma/J=0.95$, $\mstar=0.312$) the Bogoliubov depletion dominates. The exact pointer rate factor is $\Geig + J_{01}^2 \approx 5290.2$, whereas the mean-field approximation yields $\Gcoll \approx 6673.9$. The mean-field rate $\Gcoll$ therefore overestimates the exact pointer rate by $(\Gcoll - (\Geig + J_{01}^2))/(\Geig + J_{01}^2) \approx 26.2\%$ (equivalently, the exact rate is $20.7\%$ below $\Gcoll$). This substantial $\mathcal{O}(1)$ finite-$N$ discrepancy highlights precisely why the macroscopic mean-field rate formula overestimates the exact quantum decoherence in the mesoscopic regime.

The dissipator additionally leaves the localised population
difference $\Delta\rho\equiv\rho_{PP}-\rho_{RR}$ invariant at
leading order in the SCS approximation: $|P\rangle$ and $|R\rangle$
are approximate eigenstates of $\hat{J}_z$ at $\pm Nm_*/2$, hence
approximate steady states of the dephasing dissipator
(see Appendix~\ref{app:rates}, Eq.~\eqref{eq:w_diss}).
At finite $N$ where $\DeltaE>0$, the Hamiltonian then couples the
imaginary part of $\rho_{PR}$ to $\Delta\rho$, splitting the
dynamics into two channels.
The real part $\mathrm{Re}(\rho_{PR})$ decouples from this system
and undergoes pure exponential decay at $\gamma_\phi\Gcoll$
(Eq.~\eqref{eq:rate_loc}, which captures this channel).
The imaginary part $\mathrm{Im}(\rho_{PR})$ and the population
difference $\Delta\rho\equiv\rho_{PP}-\rho_{RR}$ form a coupled
2-component system whose dynamics are governed by the characteristic
polynomial $\lambda^2 + \gamma_\phi\Gcoll\lambda + \DeltaE^2=0$
(derived explicitly in Appendix~\ref{app:rates}).
In the mesoscopic secular regime where $\DeltaE\gg\gamma_\phi\Gcoll$,
the discriminant is negative: $\mathrm{Im}(\rho_{PR})$ and
$\Delta\rho$ undergo \emph{damped oscillations} at frequency $\DeltaE$
with an envelope decaying at $\gamma_\phi\Gcoll/2$~\cite{BreuerPetruccione2002}.
This envelope rate equals $\gamma_\phi\Geig$ only in the thermodynamic
limit $\eta_{\rm MF}\rightarrow2$; at finite $N$ it exceeds $\gamma_\phi\Geig$
by the factor $\eta_{\rm MF}/2$.
For the purposes of this paper, only the real-part channel is
relevant: it gives the pure exponential decay of the order-parameter
coherence at the mean-field rate $\gamma_\phi\Gcoll$.

\subsection{Rate for \boldmath$\rho_{01}$}

In the energy eigenstate basis, the secular Lindblad equation gives:
\begin{equation}
  \dot\rho_{01}\big|_{\rm diss}
  = -\gamma_\phi\Geig\,\rho_{01}.
  \label{eq:rate_eig}
\end{equation}
The secular approximation drops the off-diagonal term
$\gamma_\phi|\langle E_0|\Jz|E_1\rangle|^2$ coupling $\rho_{01}$
to $\rho_{10}$. In the interaction picture both counter-rotate at
$\omega_{01}\equiv\DeltaE/\hbar$ (the Bohr frequency of the ground
doublet), so their coupling oscillates at $2\DeltaE/\hbar$;
the secular condition is thus $(2\DeltaE/\hbar)\cdot T_2\gg1$\footnote{Here and throughout, we frequently set $\hbar=1$ for brevity, such that the energy splitting $\DeltaE$ corresponds directly to the coherent tunnelling angular frequency.}, where
$T_2\equiv(\gamma_\phi\Geig)^{-1}$ is the eigenstate coherence
decay time (formalised in Corollary~2, Section~\ref{sec:theorem}).
At the LMG benchmark ($N=370$, $\Gamma/J=0.95$,
$\gamma_\phi=0.05$~s$^{-1}$),
$(2\DeltaE/\hbar)\cdot T_2\approx18.4$~\cite{mouslopoulos2026}.

Crucially, the population-to-coherence coupling in the Redfield
tensor vanishes exactly by parity for the ground doublet
(Corollary~1, Section~\ref{sec:theorem}): $\rho_{01}$ decouples from
populations $\rho_{00}$, $\rho_{11}$ without secular approximation.
This decoupling is exact within the Markovian Lindblad dissipator
of Eq.~\eqref{eq:lindblad}; it relies on the parity proposition
$\langle E_i|\Jz|E_i\rangle=0$ (all $i$) and
$\langle E_j|\Jz^2|E_i\rangle=0$ for opposite-parity pairs
(see Section~\ref{sec:theorem}).

The ratio of the two geometric factors, Eq.~\eqref{eq:eta_def},
equals the ratio of physical rates (the $\gamma_\phi$ cancels).

It is illuminating to decompose $\eta_{\rm MF}$ into two physically
distinct factors:
\begin{equation}
  \eta_{\rm MF} = \eta_{\rm geom}\times\eta_{\rm quantum},
  \label{eq:eta_decomp}
\end{equation}
where:
\begin{align}
  \eta_{\rm geom} &\equiv 2,
  \label{eq:eta_geom}\\
  \eta_{\rm quantum} &\equiv \frac{(N\mstar/2)^2}{\Geig} \geq 1.
  \label{eq:eta_quantum}
\end{align}
$\eta_{\rm geom}=2$ is the \emph{universal geometric factor}: it
is the exact thermodynamic-limit value of $\eta_{\rm MF}$, arising
purely from the vanishing of the cross-term in the eigenstate basis
(as shown in the keyresult above), and is independent of model
parameters.
$\eta_{\rm quantum}\geq1$ is the \emph{many-body quantum
enhancement}: it exceeds unity because Bogoliubov spin-wave
depletion suppresses $\Geig$ below the mean-field reference
$(N\mstar/2)^2$ (Section~\ref{sec:degen}).
The decomposition is exact: since
$\Gcoll=(N\mstar)^2/2=2(N\mstar/2)^2$, we have
$\eta_{\rm MF} = \Gcoll/\Geig = 2(N\mstar/2)^2/\Geig
= \eta_{\rm geom}\times\eta_{\rm quantum}$.

At the benchmark $N=370$: $\eta_{\rm quantum}=1.175$, giving
$\eta_{\rm MF}=2\times1.175=2.35$.
At the peak $N\approx300$: $\eta_{\rm quantum}=1.201$, giving
$\eta_{\rm MF}=2\times1.201=2.40$.
The full $\approx21\%$ enhancement above the geometric baseline
($\eta_{\rm quantum}-1\approx0.2$) arises from many-body quantum
fluctuations near the quantum-critical crossover and reflects the
Bogoliubov suppression of $\Geig$ below the mean-field reference.

The \emph{exact} physical protection factor is separately:
$\eta_{\rm exact} = (\Geig+J_{01}^2)/\Geig = 1 + J_{01}^2/\Geig$.
At $N=370$: $\eta_{\rm exact} = (2839+2451)/2839 = 1.863$.
The difference $\eta_{\rm MF}/\eta_{\rm exact} = \Gcoll/(\Geig+J_{01}^2)
\approx1.262$ reflects the 26.2\% overestimation by the SCS mean-field
approximation (Section~\ref{sec:rates}, Approximations 1--2).

\subsection{The thermodynamic limit and the factor of 2}
\label{sec:limit}

Because the pointer states are defined exactly as $|P, R\rangle = (|E_0\rangle \pm |E_1\rangle)/\sqrt{2}$, and $\Jz^2$ is parity-even (making the cross-terms $\langle E_0|\Jz^2|E_1\rangle$ vanish exactly), the doublet average $\Geig$ corresponds precisely to the exact pointer-state second moment at any finite $N$:
\begin{equation}
  \Geig = \frac{1}{2}\langle P|\Jz^2|P\rangle +\frac{1}{2}\langle R|\Jz^2|R\rangle
  = \left(\frac{N\mstar}{2}\right)^2
    - \delta G(N,\Gamma/J)
    + \mathcal{O}(1),
  \label{eq:limit_E0}
\end{equation}
where $\delta G>0$ is the quantum-delocalisation correction of
Eq.~\eqref{eq:finite_N}: the exact eigenstate wavefunction spreads
into the classically forbidden barrier region, suppressing
$\Geig$ \emph{below} the mean-field
reference $(N\mstar/2)^2$.
Note that Eq.~\eqref{eq:limit_E0} is an exact identity at each finite $N$; the thermodynamic limit $\Geig \rightarrow (N\mstar/2)^2$ is recovered simply because the finite-$N$ quantum correction $\delta G \sim \mathcal{O}(N)$ becomes subleading to the $\mathcal{O}(N^2)$ mean-field baseline.

This should be contrasted with the mean-field spin-coherent state
$|P_{\rm SCS}\rangle$, defined as the product state (SCS) satisfying
$\langle P_{\rm SCS}|\Jz|P_{\rm SCS}\rangle = +N\mstar/2$ exactly by
construction. It is \emph{not} identical to the exact pointer state
$|P\rangle$ at finite $N$: the exact state is the half-sum of
eigenstates $(|E_0\rangle+|E_1\rangle)/\sqrt{2}$, whose mean
magnetisation satisfies $|\langle P|\Jz|P\rangle| = J_{01} \leq Nm_*/2$
(with the approach to $Nm_*/2$ controlled by the same exponentially
small corrections as Eqs.~\eqref{eq:E0}--\eqref{eq:E1}).
The SCS second moment is:
\begin{equation}
  \langle P_{\rm SCS}|\Jz^2|P_{\rm SCS}\rangle
  = \left(\frac{N\mstar}{2}\right)^2 + \frac{N(1-\mstar^2)}{4}.
  \label{eq:Pj2_scs}
\end{equation}
The spin-coherent-state variance $N(1-\mstar^2)/4$ adds a
\emph{positive} $\mathcal{O}(N)$ correction --- of \emph{opposite}
sign to the $-\delta G$ correction for the exact eigenstate.
Both corrections are subleading at $\mathcal{O}(N^{-1})$ relative to
$(N\mstar/2)^2$.
The SCS approximation $|P\rangle\approx|P_{\rm SCS}\rangle$ is used
implicitly in the localised-state rate derivation of
Section~\ref{sec:rates} (where $\langle P|\Jz|P\rangle$ is set
to $+N\mstar/2$); this introduces a relative error of
$\mathcal{O}(N^{-1})$ from Bogoliubov spin-wave depletion
(distinct from, and much larger than, the
$\mathcal{O}(S)$ vacuum overlap corrections in
Eqs.~\eqref{eq:E0}--\eqref{eq:E1}, which describe the
eigenstate-to-pointer-state relation, not the
pointer-to-SCS relation).

In the thermodynamic limit only the leading $(N\mstar/2)^2$ term
survives in both expressions, and identically for
$\langle E_1|\Jz^2|E_1\rangle$, giving
\begin{equation}
  \Geig \xrightarrow{N\rightarrow\infty}
  \frac{(N\mstar)^2}{4},
\end{equation}
and since $\Gcoll=(N\mstar)^2/2$:
\begin{equation}
  \eta_{\rm MF}(N,\Gamma/J) \xrightarrow{N\rightarrow\infty} 2.
  \label{eq:eta_limit}
\end{equation}

\begin{keyresult}[Geometric factors vs physical rates]
Equation~\eqref{eq:eta_limit} is a statement about the
\emph{geometric factors} $\Gcoll$ and $\Geig$, not about physical
dephasing rates in the thermodynamic limit.
As $N\rightarrow\infty$, $\DeltaE$ vanishes exponentially while
$\Geig\sim N^2$, so $2\DeltaE\cdot T_2\sim e^{-cN}/N^2\rightarrow0$:
the secular approximation underlying $\gamma_\phi\Geig$ breaks down.
The physically meaningful regime is the mesoscopic secular window
($N\approx250$--$430$, $\Gamma/J=0.95$), where
$2\DeltaE\cdot T_2\approx18$ and both rates are well-defined
exponential decay constants.
The degenerate-limit physics is treated analytically in
Section~\ref{sec:degen} below.
\end{keyresult}

\subsection{The degenerate limit: pointer basis and convergence of
rates}
\label{sec:degen}

As $N\rightarrow\infty$ and the doublet becomes degenerate, the
secular approximation fails. The correct treatment is to diagonalise
the Lindblad generator \emph{within the degenerate subspace}.
In the doublet $\{|E_0\rangle,|E_1\rangle\}$, the Parity Proposition
(Section~\ref{sec:theorem}) forces $\langle E_i|\Jz|E_i\rangle=0$,
so $\Jz$ is purely off-diagonal in the doublet subspace:
\begin{equation}
  \Jz\big|_{\rm doublet} = \Jol\,\sigma_x,
  \label{eq:Jz_doublet_lin}
\end{equation}
where $\Jol\equiv\langle E_0|\Jz|E_1\rangle$ and $\sigma_x$ is the
Pauli matrix in the $\{|E_0\rangle,|E_1\rangle\}$ basis.
Note that using the exact definitions $|E_0\rangle=(|P\rangle+|R\rangle)/\sqrt{2}$ and $|E_1\rangle=(|P\rangle-|R\rangle)/\sqrt{2}$, one finds exactly for all $N$:
\begin{equation*}
  \Jol = \langle E_0|\Jz|E_1\rangle
  = \tfrac{1}{2}(\langle P|+\langle R|)\Jz(|P\rangle-|R\rangle)
  = \tfrac{1}{2}\bigl(\langle P|\Jz|P\rangle - \langle R|\Jz|R\rangle\bigr).
\end{equation*}
In the thermodynamic limit, this expectation value approaches $\frac{N\mstar}{2}$,
so that $J_{01}$ is the same quantity as $\langle P|\Jz|P\rangle$ in
the thermodynamic limit.
This is not a notational accident: it reflects the fact that the
off-diagonal Lindblad matrix element in the eigenstate basis equals
the magnetisation of the pointer state, both converging to
$N\mstar/2$ as $N\rightarrow\infty$~\cite{mouslopoulos2026}.

For $\Jz^2$, the exact diagonal element within the full Hilbert space is:
\begin{equation}
  \langle E_i|\Jz^2|E_i\rangle
  = \Jol^2 + \sum_{k\geq2}|\langle E_i|\Jz|E_k\rangle|^2,
  \label{eq:Jz2_exact}
\end{equation}
where the sum over $k\geq2$ represents $\mathcal{O}(N)$ population
leakage from the doublet into the higher Dicke states via the bath.
Because $\Geig$ is the doublet average, we must account for both states: $|E_1\rangle$ sits closer to the barrier top and penetrates further into the classically forbidden region, producing significantly larger matrix elements with higher Dicke states than $|E_0\rangle$. Numerically, the average leakage accounts for $\approx 33.4\%$ at $N=100$ ($1.1\%$ for $E_0$ vs $49.8\%$ for $E_1$), $\approx 13.7\%$ at $N=370$ ($4.8\%$ for $E_0$ vs $21.0\%$ for $E_1$), and $\approx 3.4\%$ at $N=1000$ of the total $\Geig$.
To obtain a self-contained dissipator within the doublet, we restrict
to the low-energy manifold and set:
\begin{equation}
  \Jz^2\big|_{\rm doublet} = \Jol^2\,\mathbbm{1},
  \label{eq:Jz2_doublet}
\end{equation}
explicitly dropping the $k\geq2$ leakage terms.
This is physically justified because the internal doublet dephasing
($\mathcal{O}(N^2)$, set by $G_{01}$) dominates the leakage
($\mathcal{O}(N)$) by a factor $\sim N$ at large $N$; the resulting
error in the restricted Lindblad eigenvalues is $\mathcal{O}(N^{-1})$.
The doublet description is therefore valid throughout the Goldilocks
window where the gap ratio $(E_2-E_0)/\DeltaE\gg1$
(Table~\ref{tab:eta}); it is not claimed to be exact for the full
many-body Hilbert space Lindbladian.
The Lindblad dissipator restricted to the doublet is therefore:
\begin{equation}
  \mathcal{D}[\rho]\big|_{\rm doublet}
  = \gamma_\phi\Jol^2\!\left(\sigma_x\rho\sigma_x - \rho\right).
  \label{eq:doublet_diss}
\end{equation}
This is \emph{dephasing in the $\sigma_x$ eigenbasis}.
The eigenstates of $\sigma_x$ in the doublet are
$|P\rangle=(|E_0\rangle+|E_1\rangle)/\sqrt{2}$ and
$|R\rangle=(|E_0\rangle-|E_1\rangle)/\sqrt{2}$ --- the pointer
states selected by the bath~\cite{Zurek2003}.

The $4\times4$ Lindblad superoperator corresponding to
Eq.~\eqref{eq:doublet_diss} has spectrum:
\begin{equation}
  \mathrm{spec}(\mathcal{L})\big|_{\rm doublet}
  = \gamma_\phi\Jol^2\times\{0,\;0,\;-2,\;-2\}.
  \label{eq:spec}
\end{equation}
The two zero eigenvalues are the steady states \emph{within the restricted doublet}: the total population
$\tr(\rho)$ and the pointer population difference
$\rho_{PP}-\rho_{RR}$ (which equals $2\mathrm{Re}(\rho_{01})$).
The two eigenvalues $-2\gamma_\phi\Jol^2$ govern the decaying modes:
the eigenstate population difference $\rho_{00}-\rho_{11}$ and the
imaginary part of the eigenstate coherence $\mathrm{Im}(\rho_{01})$.
Together, these decaying modes constitute the full complex pointer
coherence $\rho_{PR}$, which decays strictly at $2\gamma_\phi\Jol^2$.

\begin{keyresult}[Quasi-steady state vs.\ true steady state]
Within the restricted 2-level doublet model,
$\mathrm{Re}(\rho_{01})$ is a zero eigenvalue of the doublet
Lindblad superoperator and therefore a steady state of that
restricted system.
In the \emph{full} $(N+1)$-dimensional Hilbert space, however,
the Lindblad operator $\hat{J}_z$ couples the doublet to
higher-energy Dicke states at rate $\gamma_\phi\times(\text{leakage})
\sim\gamma_\phi\mathcal{O}(N)$ (Section~\ref{sec:degen},
Eq.~\eqref{eq:Jz2_exact}).
The pure dephasing channel has no cooling: the system ultimately
thermalises to the maximally mixed state over the full Hilbert space.
$\mathrm{Re}(\rho_{01})$ is therefore a \emph{quasi-steady state}
(metastable plateau): it remains approximately constant throughout
the doublet-dephasing phase (of duration $\sim({\gamma_\phi G_{01}})^{-1}
\sim(\gamma_\phi N^2)^{-1}$, during which $\mathrm{Im}(\rho_{01})$
and $\rho_{00}-\rho_{11}$ decay), and only decays on the much longer
leakage timescale $(\gamma_\phi N)^{-1}$.
The timescale separation $G_{01}/\text{leakage}\approx7.3$ at $N=370$
(from Table~\ref{tab:eta}: $G_{01}=2839$ vs.\ average leakage $\approx388$)
ensures the doublet description remains physically well-defined for the
mesoscopic secular window.
All rate formulas in this paper are valid at intermediate times
$(\gamma_\phi N^2)^{-1}\ll t\ll(\gamma_\phi N)^{-1}$.
\end{keyresult}

Since $|P\rangle\equiv(|E_0\rangle+|E_1\rangle)/\sqrt{2}$ by definition,
$|E_0\rangle=(\Pket+\Rket)/\sqrt{2}$ is an exact algebraic identity for all $N$,
so $\Jol=\langle P|\Jz|P\rangle$ exactly.
As $N\rightarrow\infty$, $|\tilde{P}\rangle\rightarrow|P\rangle$
(correction $\mathcal{O}(S)\rightarrow0$), giving $\Jol\rightarrow N\mstar/2$, and:
\begin{equation}
  \Jol^2 \xrightarrow{N\rightarrow\infty}
  \left(\frac{N\mstar}{2}\right)^2 = \Geig.
  \label{eq:J01_limit}
\end{equation}
The doubly-degenerate nonzero decay rate therefore satisfies:
\begin{equation}
  2\gamma_\phi\Jol^2
  \xrightarrow{N\rightarrow\infty}
  2\gamma_\phi\Geig = \gamma_\phi\Gcoll.
  \label{eq:rate_convergence}
\end{equation}
The decaying components of both coherences are governed by
$\gamma_\phi\Gcoll$ in the thermodynamic limit.
Numerical verification is shown in Figure~\ref{fig:convergence}.

\begin{keyresult}[Three-regime structure]
The physical content of $\eta_{\rm MF}=\Gcoll/\Geig$ is organised into three
regimes:

\smallskip
\noindent\textbf{Regime~1 --- Small $N$ ($N\lesssim100$):}
The double-well is not yet established ($Nm_*\lesssim30$).
$\eta_{\rm MF}<2$ and can fall below $1$.
Neither rate formula gives a well-defined exponential decay constant:
$\Gcoll=(N\mstar)^2/2$ overestimates the actual order parameter
because the mean-field value $\mstar$ is not yet established at finite
$N$, while $\Geig$ loses meaning as the doublet is not spectrally
isolated from higher states (gap ratio $\lesssim3$, Table~\ref{tab:eta})
and the SCS approximation has relative error $\gtrsim9\%$.

\smallskip
\noindent\textbf{Regime~2 --- Mesoscopic secular window
($N\approx250$--$430$):}
The secular condition $2\DeltaE\cdot T_2\gg1$ is satisfied
($\approx18.4$ at $N=370$) and the doublet is well-isolated from
higher states (gap ratio $(E_2-E_0)/\DeltaE\approx11.4$).
Two coherences with \emph{genuinely different} exponential decay
rates:
\begin{equation*}
  \dot\rho_{01}\big|_{\rm diss}=-\gamma_\phi\Geig\,\rho_{01},
  \qquad
  \dot\rho_{PR}\big|_{\rm diss}\simeq-\gamma_\phi\Gcoll\,\rho_{PR},
  \qquad\eta_{\rm MF}\approx2.35\text{--}2.40.
\end{equation*}
This is the protected quantum regime. It encompasses the thermal
``Goldilocks'' zone ($N\approx250$--$370$) defined in the companion
paper~\cite{mouslopoulos2026} --- demonstrating that the dynamical
protection of the secular approximation outlasts the strict thermal
bounds, allowing quantum technologies to harvest a coherence lifetime
extended by a factor of $\eta_{\rm exact}\approx1.86$ (exact physical protection)
or a factor of $\eta_{\rm MF}\approx2.35$--$2.40$ (advantage over classical mean-field).

\smallskip
\noindent\textbf{Regime~3 --- Thermodynamic limit
($N\rightarrow\infty$):}
The secular approximation fails ($\DeltaE\rightarrow0$, doublet
degenerate). The Lindblad generator in the doublet is
Eq.~\eqref{eq:doublet_diss}, with the doubly-degenerate nonzero rate
$2\gamma_\phi\Jol^2\rightarrow\gamma_\phi\Gcoll$.
The pointer coherence $\rho_{PR}$ and $\mathrm{Im}(\rho_{01})$ decay
at $\gamma_\phi\Gcoll$; $\mathrm{Re}(\rho_{01})$ approaches a quasi-steady plateau within the doublet (true steady state only in the doublet approximation; it decays on the longer leakage timescale $\sim(\gamma_\phi N)^{-1}$ in the full Hilbert space).
\emph{The two pictures converge.}
The factor of $\eta_{\rm MF}$ is a \emph{mesoscopic quantum effect}, not a
universal constant.
\end{keyresult}

\noindent
The factor of $2$ has a precise physical origin.
$\Gcoll$ is \emph{half} the squared distance between macroscopic
magnetisations:
$\Gcoll=(\langle P|\Jz|P\rangle-\langle R|\Jz|R\rangle)^2/2
=(N\mstar)^2/2=2(N\mstar/2)^2$.
Note that $\langle P|\Jz^2|P\rangle\rightarrow(N\mstar/2)^2$
and $\langle E_0|\Jz^2|E_0\rangle\rightarrow(N\mstar/2)^2$ converge
to the same value individually --- their ratio approaches $1$, not $2$.
For the eigenstate, this $\mathcal{O}(N^2)$ variance is not standard
statistical noise, but the macroscopic quantum uncertainty of a
superposition between $+N\mstar/2$ and $-N\mstar/2$.
The factor of $2$ arises because $\Gcoll$ is \emph{twice} this
single-state variance: it measures the squared distance between two
states, not the variance of one.
In the localised basis, the cross-term
$\langle P|\Jz|P\rangle\langle R|\Jz|R\rangle=-(N\mstar/2)^2$
doubles the rate; in the eigenstate basis, the analogous cross-term
$\langle E_0|\Jz|E_0\rangle\langle E_1|\Jz|E_1\rangle=0$ exactly
by parity, for all $N$.

\section{Finite-\boldmath{$N$} Amplification}
\label{sec:finite}

At finite $N$ the mean-field ratio $\eta_{\rm MF}$ has a richer structure.
For very small $N$, where the double-well structure is not yet
well-developed and the eigenstates are delocalised ($N\mstar\ll1$),
the eigenstate second moment $\langle E_i|\Jz^2|E_i\rangle$ exceeds
$(N\mstar/2)^2$: $\eta_{\rm MF}<1$.
As $N$ increases into the ordered phase, $\eta_{\rm MF}$ passes through $1$
(around $N\approx25$--$30$ at $\Gamma/J=0.95$), rises to a maximum
$\eta_{\rm MF,peak}\approx2.40$ near $N\approx300$, then decreases
monotonically toward $\eta_{\rm MF}=2$ as $N\rightarrow\infty$.

The amplification above $2$ at finite $N$ has a precise origin:
$\langle E_i|\Jz^2|E_i\rangle$ is suppressed below $(N\mstar/2)^2$
by quantum fluctuations as the eigenstate wavefunction spreads into
the classically forbidden region near $m_z=0$:
\begin{equation}
  \Geig = \left(\frac{N\mstar}{2}\right)^2 - \delta G(N,\Gamma/J),
  \quad
  \delta G \sim \mathcal{O}(N).
  \label{eq:finite_N}
\end{equation}
Hence $\eta_{\rm MF}=2+\mathcal{O}(N^{-1})$, with the correction largest near
the crossover $\DeltaE\approx k_BT$~\cite{mouslopoulos2026},
where the wavefunction spread is maximum.

\section{The LMG Model: Exact Numerics}
\label{sec:lmg}

The Lipkin-Meshkov-Glick Hamiltonian~\cite{Lipkin1965,Botet1982}
is the $N$-spin realisation of Eq.~\eqref{eq:H}, exactly solvable
by diagonalisation of the $(N+1)$-dimensional symmetric Dicke sector.

\subsection{The ratio \boldmath$\eta_{\rm MF}(N,\Gamma/J)$}

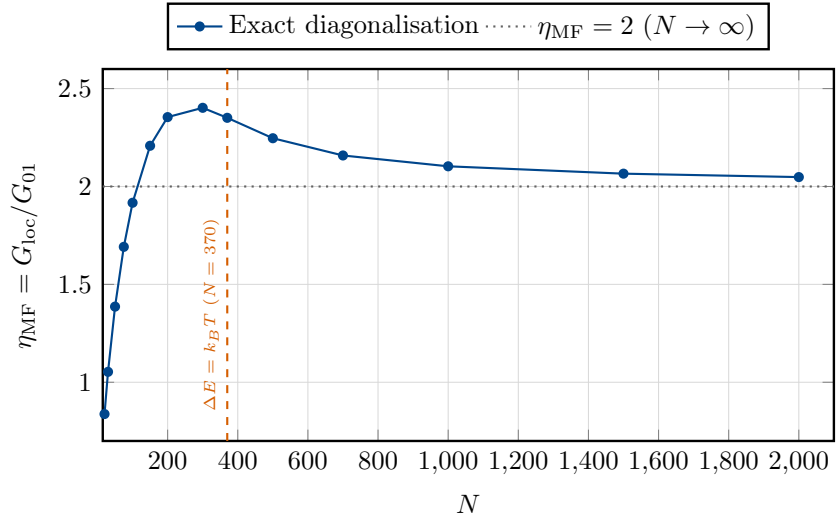
\begin{figure}[H]
\centering
\begin{tikzpicture}
\begin{axis}[
  qplot,
  width=0.75\linewidth, height=6.5cm,
  xlabel={$N$},
  ylabel={$\eta_{\rm MF} = G_{\rm loc}/G_{01}$},
  xmin=15, xmax=2100,
  ymin=0.7, ymax=2.6,
  legend style={at={(0.5,1.05)}, anchor=south, font=\small,
                legend columns=2},
]
\addplot[color=qblue, mark=*, mark size=1.5pt, thick, solid]
  coordinates {
  (20,  0.8376)(30,  1.0532)(50,  1.3863)(75,  1.6916)
  (100, 1.9169)(150, 2.2080)(200, 2.3544)(300, 2.4019)
  (370, 2.3508)(500, 2.2465)(700, 2.1584)(1000,2.1031)
  (1500,2.0655)(2000,2.0480)
};
\addplot[color=qgray, dotted, thick, domain=15:2100] {2.0};
\addlegendentry{Exact diagonalisation}
\addlegendentry{$\eta_{\rm MF}=2$ ($N\rightarrow\infty$)}
\draw[dashed,qred,thick] (axis cs:370,0.7) -- (axis cs:370,2.6);
\node[font=\tiny,qred,rotate=90,anchor=south west] at
  (axis cs:378,0.80) {$\DeltaE=k_BT$ ($N=370$)};
\end{axis}
\end{tikzpicture}
\caption{Ratio $\eta_{\rm MF}=G_{\rm loc}/G_{01}$ versus $N$ at $\Gamma/J=0.95$,
from exact diagonalisation. $\eta_{\rm MF}<1$ for small $N$ (double-well not
yet developed); peaks at $\eta_{\rm MF}\approx2.40$ near $N=300$; decreases
toward the geometric lower bound $\eta_{\rm MF}=2$ (dotted). The crossover
$\DeltaE=k_BT$ (dashed red) sits at $N=370$.}
\label{fig:eta_N}
\end{figure}

Figure~\ref{fig:eta_N} shows $\eta_{\rm MF}$ versus $N$ at $\Gamma/J=0.95$.

Table~\ref{tab:eta} lists the key values. The $N\rightarrow\infty$
row gives the analytic thermodynamic-limit result.

\begin{table}[H]
\centering
\caption{Dimensionless geometric factors $G_{\rm loc}\equiv(N m_*)^2/2$
and $G_{01}\equiv\tfrac{1}{2}(\langle E_0|\Jz^2|E_0\rangle
+\langle E_1|\Jz^2|E_1\rangle)$ at $\Gamma/J=0.95$ ($m_*=0.312$).
The mean-field dephasing rate is $\gamma_\phi G_{\rm loc}$; the exact eigenstate rate is $\gamma_\phi G_{01}$.
The ratio $\eta_{\rm MF}$ is independent of $\gamma_\phi$.
The gap ratio $(E_2-E_0)/\DeltaE$ quantifies doublet isolation from
higher states. The secular parameter $2\DeltaE\cdot T_2$ is evaluated
at the benchmark $\gamma_\phi=0.05$~s$^{-1}$~\cite{mouslopoulos2026};
values $\gg1$ confirm
the secular approximation; values $\lesssim3$ indicate its breakdown.
All entries are computed from full-precision exact diagonalisation
and rounded independently; small inconsistencies in the last
displayed digit (e.g.\ $488/254=1.921$ vs $\eta_{\rm MF}=1.917$ at $N=100$)
reflect this independent rounding.}
\label{tab:eta}
\begin{tabular}{ccccccc}
\toprule
$N$ & $\DeltaE/k_BT$ & $G_{\rm loc}$ & $G_{01}$ & $\eta_{\rm MF}$
    & $(E_2{-}E_0)/\DeltaE$ & $2\DeltaE T_2$ \\
\midrule
 100  & 7.33  &    488  &    254 & 1.917 &    3.0  & 1510  \\
 200  & 3.63  &   1950  &    828 & 2.354 &    4.1  &  230  \\
 300  & 1.75  &   4388  &   1827 & 2.402 &    6.9  &   50  \\
 370  & 1.00  &   6674  &   2839 & 2.351 &   11.4  &   18  \\
 500  & 0.32  &  12188  &   5425 & 2.246 &   37.3  &    3  \\
1000  & $\ll1$&  48750  &  23180 & 2.103 & $>10^3$ & $\ll1$ \\
2000  & $\ll1$& 195000  &  95214 & 2.048 & $>10^3$ & $\ll1$ \\
$\infty$ & $0$ & $N^2m_*^2/2$ & $N^2m_*^2/4$ & $2$ & $\infty$ & $0$ \\
\bottomrule
\end{tabular}
\end{table}

\subsection{Dependence on \boldmath$\Gamma/J$}

\begin{figure}[H]
\centering
\begin{tikzpicture}
\begin{axis}[
  qplot,
  width=0.75\linewidth, height=8cm,
  xlabel={$\Gamma/J$},
  ylabel={$\eta_{\rm MF} = G_{\rm loc}/G_{01}$},
  xmin=0, xmax=1.0,
  ymin=0.4, ymax=2.6,
  legend style={at={(0.5,1.05)}, anchor=south, font=\small,
                legend columns=2},
]
\addplot[color=qblue, mark=square*, mark size=1.5pt, thick, solid]
  coordinates {
  (0.10,2.0001)(0.20,2.0002)(0.30,2.0006)(0.40,2.0012)
  (0.50,2.0022)(0.60,2.0043)(0.70,2.0086)(0.80,2.0207)
  (0.85,2.0371)(0.90,2.0828)(0.92,2.1308)(0.94,2.2459)
  (0.95,2.3508)(0.96,2.4244)(0.97,2.3092)(0.98,1.8827)
  (0.99,1.1101)(0.995,0.5961)
};
\addplot[color=qgray, dotted, thick, domain=0:1.0] {2.0};
\addlegendentry{$N=370$, exact diag.}
\addlegendentry{$\eta_{\rm MF}=2$ ($N\rightarrow\infty$)}
\draw[->,thick,qgray] (axis cs:0.95,2.3508) -- (axis cs:0.70,2.20)
  node[left,font=\scriptsize,align=right]
  {$\Gamma/J{=}0.95$\\$\eta_{\rm MF}{=}2.35$};
\draw[->,thick,qorange] (axis cs:0.96,2.4244) -- (axis cs:0.80,2.40)
  node[left,font=\scriptsize,align=right,qorange]
  {peak: $\Gamma/J{=}0.96$\\$\eta_{\rm MF}{=}2.42$};
\end{axis}
\end{tikzpicture}
\caption{Ratio $\eta_{\rm MF}$ versus $\Gamma/J$ at $N=370$.
$\eta_{\rm MF}\rightarrow2$ deep in the ordered phase ($\Gamma/J\rightarrow0$);
maximum $\eta_{\rm MF}\approx2.42$ at $\Gamma/J\approx0.96$ (orange arrow);
benchmark $\Gamma/J=0.95$ marked (grey arrow).
$\eta_{\rm MF}$ falls sharply through $1$ and approaches $0$ near criticality.}
\label{fig:eta_gamma}
\end{figure}
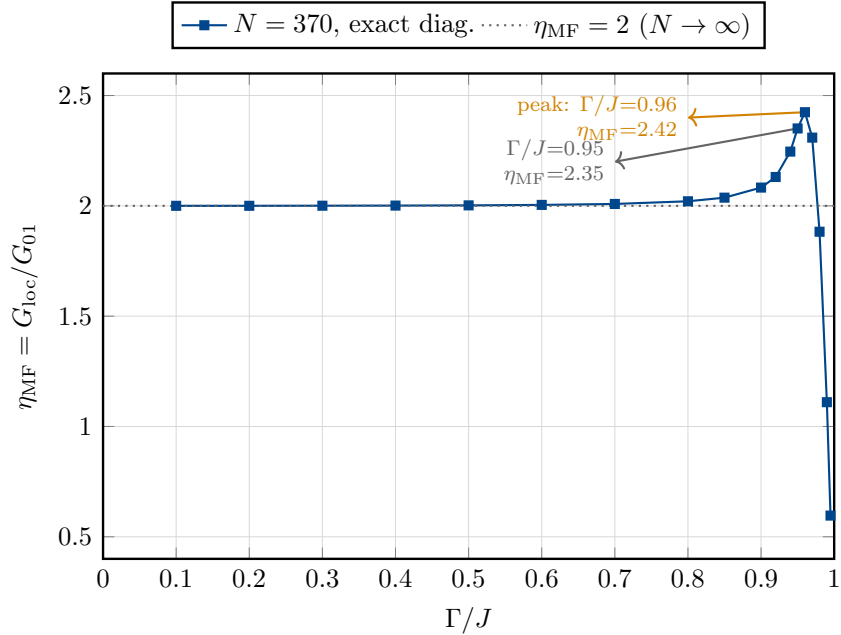

Figure~\ref{fig:eta_gamma} shows $\eta_{\rm MF}$ versus $\Gamma/J$ at $N=370$.

\subsection{Convergence: geometric factors and physical rates}

Fitting $\eta_{\rm MF}(N)=2+c/N$ at $\Gamma/J=0.95$ gives
$c\approx123$ at $N=500$, decreasing to $c\approx103$ at $N=1000$
and $c\approx96$ at $N=2000$,
confirming the $\mathcal{O}(N^{-1})$ approach of the geometric ratio
to $2$.
The factor of $2$ in $\eta_{\rm MF}$ is not a finite-$N$ artefact.

However, the geometric convergence $\eta_{\rm MF}\rightarrow2$ is not the same
as the convergence of physical rates.
As established in Section~\ref{sec:degen}, the full Lindblad
superoperator in the degenerate doublet has a doubly-degenerate
nonzero eigenvalue $2\gamma_\phi\Jol^2$, which converges to
$\gamma_\phi\Gcoll$ from below.
Simultaneously, the secular rate $\gamma_\phi\Geig$ converges to
$\gamma_\phi\Gcoll/2$.
Both physical rates therefore converge to $\gamma_\phi\Gcoll$;
$\eta_{\rm MF}\rightarrow2$ is the geometric expression of this convergence.

\begin{figure}[H]
\centering
\begin{tikzpicture}
\begin{axis}[
  qplot,
  width=0.75\linewidth, height=7cm,
  xlabel={$N$},
  ylabel={Rate factor\,/\,$(N\mstar/2)^2$},
  xmin=15, xmax=2100,
  ymin=0.6, ymax=2.5,
  legend style={at={(0.5,-0.30)}, anchor=north, font=\small,
                legend cell align=left},
]
\addplot[color=qred, dotted, very thick, domain=15:2100] {2.0};
\addplot[color=qblue, mark=*, mark size=1.5pt, thick, solid]
  coordinates {
  (20, 3.0047)(30, 2.3895)(50, 1.8364)(75, 1.5369)
  (100,1.3891)(150,1.2707)(200,1.2585)(250,1.2979)
  (300,1.3638)(370,1.4692)(500,1.6284)(700,1.7554)
  (1000,1.8372)(1500,1.8948)(2000,1.9222)
};
\addplot[color=qorange, mark=square*, mark size=1.5pt, thick, dashed]
  coordinates {
  (20, 2.3877)(30, 1.8989)(50, 1.4427)(75, 1.1823)
  (100,1.0433)(150,0.9058)(200,0.8495)(250,0.8310)
  (300,0.8327)(370,0.8508)(500,0.8903)(700,0.9266)
  (1000,0.9510)(1500,0.9683)(2000,0.9766)
};
\addplot[color=qgray, dotted, thick, domain=15:2100] {1.0};
\addlegendentry{$G_{\rm loc}/(N\mstar/2)^2 = 2$}
\addlegendentry{$2J_{01}^2/(N\mstar/2)^2$ (restricted doublet eigenvalue)}
\addlegendentry{$G_{01}/(N\mstar/2)^2$ (secular rate)}
\addlegendentry{$(N\mstar/2)^2/(N\mstar/2)^2=1$}
\addplot[fill=qgreen,fill opacity=0.10,draw=none] coordinates
  {(250,0.6)(250,2.5)(430,2.5)(430,0.6)} \closedcycle;
\node[font=\scriptsize,qgreen!70!black,rotate=90] at (axis cs:270,1.35)
  {Secular window};
\draw[dashed,qred,thick] (axis cs:430,0.6) -- (axis cs:430,2.5);
\node[font=\scriptsize,qblue,anchor=west] at (axis cs:1600,1.96)
  {$\rightarrow 2$};
\node[font=\scriptsize,qorange,anchor=west] at (axis cs:1600,0.98)
  {$\rightarrow 1$};
\end{axis}
\end{tikzpicture}
\caption{Three decoherence rate factors normalised by $(N\mstar/2)^2$,
the single-state second moment in the thermodynamic limit.
\textit{Red dotted} (flat at $2$): $G_{\rm loc}/(N\mstar/2)^2=2$,
the localised-state rate --- constant by definition.
\textit{Blue solid}: $2J_{01}^2/(N\mstar/2)^2$, the nonzero
eigenvalue of the restricted doublet Lindblad superoperator
(Eq.~\eqref{eq:spec}); \emph{converges to $2$}, i.e.\ merges with
$\gamma_\phi G_{\rm loc}$, as $N\rightarrow\infty$, confirming
Eq.~\eqref{eq:rate_convergence}.
\textit{Orange dashed}: $G_{01}/(N\mstar/2)^2=2/\eta_{\rm MF}$, the secular
rate; converges to $1$, i.e.\ to $\gamma_\phi G_{\rm loc}/2$ --- does
\emph{not} merge with the other two curves.
The factor $\eta_{\rm MF}=G_{\rm loc}/G_{01}\rightarrow2$ is the ratio between
the red and orange asymptotes.
The shaded region is the mesoscopic secular window
($2\DeltaE\cdot T_2\gg1$), the only regime where both rates are
simultaneously well-defined exponential decay constants.
This dynamically protected window comfortably encompasses the narrower
thermal ``Goldilocks'' zone ($N\approx250$--$370$,
$\DeltaE \gtrsim k_B T$) required for experimental
protocols~\cite{mouslopoulos2026}.}
\label{fig:convergence}
\end{figure}
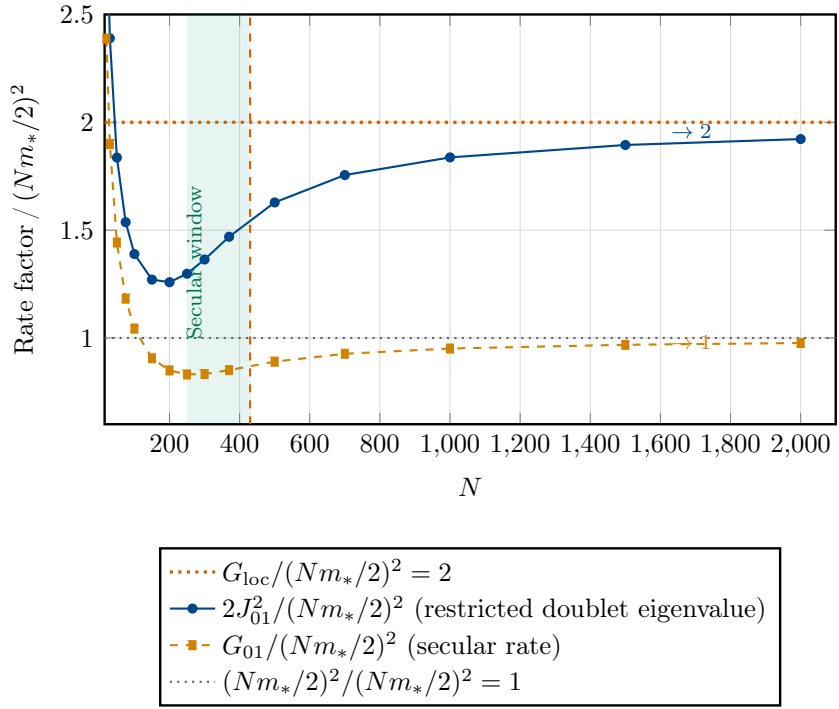

Figure~\ref{fig:convergence} shows all three rates normalised by the
common scale $(N\mstar/2)^2$, making the convergence visually
explicit.

\section{Parity Proposition}
\label{sec:theorem}

The vanishing of the cross-term in the eigenstate basis is an
exact consequence of the $\mathbb{Z}_2$ symmetry.

\smallskip
\noindent\textbf{Proposition.}
\textit{Let $[\hat{H},\hP]=0$ and $\hP\hat{L}\hP^\dagger=-\hat{L}$,
where $\hP$ is a unitary operator ($\hP^\dagger\hP=\mathbbm{1}$).
Then for any eigenstate $|E_i\rangle$ of $\hat{H}$ with
$\hP|E_i\rangle=p_i|E_i\rangle$:
$\langle E_i|\hat{L}|E_i\rangle=0$.}

\smallskip
\noindent\textit{Proof.}
Inserting $\hP^\dagger\hP=\mathbbm{1}$ on both sides of $\hat{L}$:
\begin{align}
  \langle E_i|\hat{L}|E_i\rangle
  &= \langle E_i|\hP^\dagger
       \underbrace{(\hP\hat{L}\hP^\dagger)}_{=-\hat{L}}
     \hP|E_i\rangle \nonumber\\
  &= p_i^*\cdot(-1)\cdot p_i\;
     \langle E_i|\hat{L}|E_i\rangle.
\end{align}
Since $\hP$ is unitary, $|p_i|^2=1$, so $p_i^*p_i=1$, and therefore
$\langle E_i|\hat{L}|E_i\rangle = -\langle E_i|\hat{L}|E_i\rangle$,
hence $\langle E_i|\hat{L}|E_i\rangle=0$. $\square$

\smallskip
\noindent\textit{Note on the LMG implementation.}
With $\hP=e^{i\pi\Jx}$, the operator is unitary for all $N$ and
satisfies $e^{i\pi\Jx}\Jz e^{-i\pi\Jx}=-\Jz$.
For even $N$, $\hP^2=\mathbbm{1}$ and $\hP$ is Hermitian with
eigenvalues $p_i=\pm1$.
For odd $N$, $\hP^2=-\mathbbm{1}$ and the eigenvalues are $p_i=\pm i$;
the Proposition still holds since only $|p_i|^2=1$ is required.
All numerical examples in this paper use even $N$, for which
$\hP$ is Hermitian. The alternative choice $\hP=\prod_j\hat\sigma_j^x$
is Hermitian and satisfies $\hP^2=\mathbbm{1}$ for all $N$.
The commutation $[\hat{H},\hP]=0$ holds for the zero longitudinal field Hamiltonian
of Eq.~\eqref{eq:H} and can be verified directly:
$e^{i\pi\Jx}(-\frac{2J}{N}\Jz^2 - 2\Gamma\Jx)e^{-i\pi\Jx}
= -\frac{2J}{N}\Jz^2 - 2\Gamma\Jx = \hat{H}$,
using $e^{i\pi\Jx}\Jz e^{-i\pi\Jx}=-\Jz$ and
$e^{i\pi\Jx}\Jx e^{-i\pi\Jx}=+\Jx$.

\smallskip
\noindent\textbf{Corollary 1}
(Population-coherence decoupling for opposite-parity pairs).
For $i\neq j$ with $p_i^*p_j\neq1$ (opposite parity eigenvalues),
the Redfield tensor element coupling the population $\rho_{ii}$ to the coherence $\rho_{ij}$ evaluates to:
\begin{equation}
  R_{(ii),(ij)} = \gamma_\phi \langle E_i|\Jz|E_i\rangle \langle E_j|\Jz|E_i\rangle 
                - \frac{\gamma_\phi}{2} \langle E_j|\Jz^2|E_i\rangle.
\end{equation}
Within the Markovian Lindblad dissipator used here, $R_{(ii),(ij)}=0$ exactly: 
populations and coherences between states of opposite parity decouple without 
secular approximation.
This decoupling is exact for the flat-spectrum Lindblad form of
Eq.~\eqref{eq:lindblad}; it need not hold for a general
microscopic Redfield tensor with frequency-dependent spectral density.
The first term in $R_{(ii),(ij)}$ vanishes by the Proposition
($\langle E_i|\Jz|E_i\rangle=0$); the second term
$\langle E_j|\Jz^2|E_i\rangle=0$ because $\Jz^2$ is parity-even
and cannot connect states of opposite parity.
For same-parity pairs ($p_i=p_j$), $\langle E_j|\Jz^2|E_i\rangle\neq0$
in general and the decoupling does not hold.
The ground doublet ($p_0=+1$, $p_1=-1$) is the opposite-parity case:
the decoupling of $\rho_{01}$ from populations is exact.

\smallskip
\noindent\textbf{Corollary 2}
(Secular approximation for $\rho_{01}$).
The secular approximation for $\rho_{01}$ drops the term
$\gamma_\phi|\langle E_0|\Jz|E_1\rangle|^2$ coupling $\rho_{01}$
to $\rho_{10}$. In the interaction picture, this coupling oscillates
at $2\DeltaE/\hbar$ (the counter-rotation of $\rho_{01}$ and
$\rho_{10}$); the secular condition is $(2\DeltaE/\hbar)\cdot T_2\gg1$,
where $T_2\equiv(\gamma_\phi\Geig)^{-1}$ is as defined in
Section~\ref{sec:rates}.
At the LMG benchmark,
$(2\DeltaE/\hbar)\cdot T_2\approx18.4$~\cite{mouslopoulos2026}.
This term does not vanish by parity: $\Jz$ is parity-odd and
$\langle E_0|\Jz|E_1\rangle\neq0$.

The proposition holds for any system with $[\hat{H},\hP]=0$ and
$\hP\hat{L}\hP^\dagger=-\hat{L}$: spin-boson models in the
broken-symmetry phase, the Dicke model in the superradiant phase,
molecular aggregates with H- or J-type symmetry, two-mode
BECs~\cite{Trenkwalder2016}.

\section{Implications}
\label{sec:implications}

\subsection{Quantum metrology and spin squeezing}

The quantum Fisher information $F_Q[\rho,\Jz]$ and the Wineland
spin-squeezing parameter $\xi^2$ are sensitive to the eigenstate
coherence $\rho_{01}$, not to $\rho_{PR}$, when the system dynamics
are effectively restricted to the ground-state doublet
(as in the mesoscopic window where the gap to $|E_2\rangle$
is $\approx11.4\times\DeltaE$~\cite{mouslopoulos2026}).
In general, QFI and squeezing depend on the full covariance matrix
of the collective spins; the doublet restriction is the regime in
which the present results apply directly.
Consequently, quantum sensors operating in this mesoscopic window are
affected by the slower eigenstate rate $\gamma_\phi\Geig$
rather than the exact pointer-state rate $\gamma_\phi(\Geig+J_{01}^2)$,
giving a coherence lifetime longer by a factor of
$\eta_{\rm exact}\in[1,1.86]$ compared to the pointer basis.
Against the classical mean-field estimate $\gamma_\phi\Gcoll$, the
advantage is the larger factor $\eta_{\rm MF}\in[2,2.42]$,
provided the dynamics remain effectively confined to the ground-state
doublet.
This protection is operationally defined: for protocols that
initialise, manipulate, and measure within the energy eigenbasis,
the relevant decoherence rate is $\gamma_\phi\Geig$, which is slower
than the localised-state rate $\gamma_\phi\Gcoll$ by the factor
$\eta_{\rm MF}$. This is not merely a choice of basis --- it reflects the
physical suppression of dephasing channels by the $\mathbb{Z}_2$
parity symmetry (Proposition, Section~\ref{sec:theorem}), which
forces the diagonal matrix elements of the Lindblad jump operator, $\langle E_i|\Jz|E_i\rangle=0$
exactly, eliminating the cross-term that doubles the localised rate.
The correct physical decoherence rate is $\gamma_\phi\Geig$,
with $\Geig$ given in Eq.~\eqref{eq:Geig_def}.

\subsection{Leggett-Garg inequality tests}

In the companion paper~\cite{mouslopoulos2026}, the Leggett-Garg
correlator $K_3$ exhibits a robust violation $K_3\approx1.32$ at
$\gamma_\phi\lesssim0.289$~s$^{-1}$ ($N=370$, $\Gamma/J=0.95$).
This threshold is set by $\Geig$, not $\Gcoll$.
Using the mean-field estimate $\gamma_\phi\Gcoll$ in place of
$\gamma_\phi\Geig$ would predict a Level~A threshold of
$\gamma_\phi\lesssim0.050$~s$^{-1}$ --- a factor of $\eta_{\rm MF}=2.35\times$
below the correct two-level eigenstate threshold of $0.117$~s$^{-1}$
(Level~B, this paper).
The further factor from multi-level Lindblad effects (Level~B to
Level~C in Ref.~\cite{mouslopoulos2026}) brings the total up to $5.8\times$ enhancement compared to mean-field estimates:
$5.8\times = 2.35\times$ (parity cross-term, this paper)
$\times\,2.47\times$ (additional odd-parity Dicke states).
These factors are computed from full-precision values
($\gamma_\phi^{\rm (B)}=0.1173$~s$^{-1}$,
$\gamma_\phi^{\rm (A)}=0.0499$~s$^{-1}$) and rounded independently;
multiplying the rounded figures introduces a residual discrepancy of
$\lesssim0.5\%$.
These contributions are physically distinct and independently
computable.

\subsection{General collective spin systems}

The result applies to anisotropic collective spin models (such as the infinite-range Ising or anisotropic Heisenberg models), the Dicke model
in the superradiant phase, two-mode BECs in the broken-symmetry
phase~\cite{Trenkwalder2016}, and molecular aggregates with
site-exchange symmetry. In each case the exact pointer-state rate exceeds the eigenstate rate
by a factor of $\eta_{\rm exact}\in[1,\eta_{\rm exact,peak}]$, and the mean-field
estimate exceeds it by a factor of $\eta_{\rm MF}\in[2,\eta_{\rm MF,peak}]$,
where the thermodynamic lower bound $\eta_{\rm MF}=2$ is universal
(Section~\ref{sec:limit}) but the finite-$N$ peaks are model-dependent
and must be computed from the specific instanton action and crossover
dynamics of the system in question.

\section{Discussion}
\label{sec:discussion}

The central point is simple: \emph{if your protocol initialises and
measures in the energy eigenstate basis, use $\gamma_\phi\Geig$.
If your protocol uses localised order-parameter states, use
$\gamma_\phi\Gcoll$. They are not the same.}

The factor of $2$ between them in the thermodynamic limit is the
ratio of half the squared distance between macroscopic first
moments, $(N\mstar)^2/2$, to the eigenstate variance $(N\mstar/2)^2$.
It is not the ratio of two second moments (both
$\langle P|\Jz^2|P\rangle$ and $\langle E_0|\Jz^2|E_0\rangle$
converge to $(N\mstar/2)^2$ individually; their ratio approaches
$1$); it is the factor by which half the squared distance exceeds the
single-state variance.
The cross-term $\langle P|\Jz|P\rangle\langle R|\Jz|R\rangle
=-(N\mstar/2)^2$ in the localised basis doubles the rate; the
analogous term vanishes exactly in the eigenstate basis by parity.

The basis dependence of the rates --- how can the same physical
system have two different decoherence rates? --- is clarified by
the three-regime structure of Section~\ref{sec:degen}: the two
rates correspond to different observables (the order-parameter
coherence $\rho_{PR}$ and the energy-eigenstate coherence
$\rho_{01}$) and are not in contradiction.
In the thermodynamic limit, the degenerate doublet analysis shows
that the decaying components of both coherences are governed by
$\gamma_\phi\Gcoll$: $\mathrm{Re}(\rho_{01})$ becomes a quasi-steady state
and the remainder decays at the doubly-degenerate nonzero Lindblad eigenvalue
$2\gamma_\phi\Jol^2\rightarrow\gamma_\phi\Gcoll$.
The rate difference exists only in the mesoscopic secular window
where $\Delta E > 0$ renders the secular approximation valid and the
two coherences genuinely distinct.

This is precisely where quantum protocols sensitive to
$\rho_{01}$ are most affected. For such protocols --- those
initialised, manipulated, and read out within the energy
eigenbasis, with dynamics effectively confined to the ground-state
doublet --- the mesoscopic secular window provides an operational
coherence lifetime longer by a factor of $\eta_{\rm exact}\lesssim2$ (specifically $\approx1.86$ at the benchmark)
compared to the exact pointer-state decay rate, or up to
a factor of $\eta_{\rm MF}\approx2.42$ compared to the classical mean-field
estimate $\gamma_\phi\Gcoll$.

The structure $\langle\Jz\rangle_{|P\rangle}\xrightarrow{N\rightarrow\infty}Nm_*/2$ (which diverges
as $N\rightarrow\infty$) while $\langle\Jz\rangle_{|E_0\rangle}=0$
exactly by parity reflects
the fundamental difference between statistical variance and
macroscopic quantum uncertainty. In the localised state, the variance
is standard statistical noise $\mathcal{O}(N)$; in the symmetric
eigenstate, the $\mathcal{O}(N^2)$ variance is the signature of a
macroscopic Schr\"{o}dinger cat state. The factor of $2$ isolates
this distinction.
The mesoscopic secular window is the finite-system analogue of the
regime where spontaneous symmetry breaking produces physical
observables distinct from those of the symmetric phase.

At finite $N$ near the crossover, the mean-field ratio reaches
$\eta_{\rm MF}\approx2.35$--$2.42$ and the exact physical protection
factor $\eta_{\rm exact}\approx1.86$, precisely where quantum protocols
are most sensitive. The Proposition of Section~\ref{sec:theorem} is
elementary; its consequences for decoherence calculations in
symmetry-broken phases are not.

\section*{Acknowledgements}

This work builds on the framework of Ref.~\cite{mouslopoulos2026}.

\appendix
\section{Numerical Methods}
\label{app:numerics}

All matrix elements are computed by exact diagonalisation of the LMG
Hamiltonian in the symmetric Dicke sector $\{|J_t,m\rangle:
m=-J_t,\ldots,+J_t\}$, $J_t=N/2$.
The $(N+1)\times(N+1)$ Hamiltonian is tridiagonal with diagonal
elements $-2Jm^2/N$ and off-diagonal elements
$-\Gamma\sqrt{J_t(J_t+1)-m(m+1)}$.
Diagonalisation uses \texttt{scipy.linalg.eigh} (LAPACK
\texttt{dsyevd} backend) in double-precision floating point
(\texttt{float64}).
Matrix elements $\langle E_i|\Jz^2|E_i\rangle$ and
$\langle E_i|\Jz|E_j\rangle$ are extracted directly from the
eigenvector coefficients in the Dicke basis.
Numerical convergence was validated by verifying:
(i)~the completeness relation
$\sum_k|\langle E_i|\Jz|E_k\rangle|^2=\langle E_i|\Jz^2|E_i\rangle$
(to machine precision, $\sim10^{-12}$, for all $N$ in Table~\ref{tab:eta});
(ii)~the parity symmetry $\langle E_i|\Jz|E_i\rangle=0$ (satisfied to
$\sim10^{-14}$ for all even-parity eigenstates);
(iii)~the $\eta_{\rm MF}(N)=2+c/N$ scaling, confirmed by the fit
$c\approx103$ at $N=1000$ and $c\approx96$ at $N=2000$.
For $N=2000$, the $(2001)\times(2001)$ matrix is diagonalised in
$\sim10$~s on a standard workstation; no convergence issues were
observed.
The companion paper~\cite{mouslopoulos2026} provides the full
self-tested simulation code at
\url{https://arxiv.org/abs/2604.18638}.

\section{Derivation of the Two-Channel Decay Structure}
\label{app:rates}

This appendix provides a self-contained derivation of the decay
structure of $\rho_{PR}$ in the localised basis, resolving the two
coupled channels identified in Section~\ref{sec:rates}.

\subsection{Full projection of the Lindblad equation}

We restrict the dynamics to the 2D projected density matrix within the
pointer-state doublet $\{|P\rangle,|R\rangle\}$. We explicitly parameterise
the complex coherence as $\rho_{PR} = u + iv$, where:
\begin{equation}
  u \equiv \mathrm{Re}(\rho_{PR}),
  \quad
  v \equiv \mathrm{Im}(\rho_{PR}),
  \quad
  w \equiv \rho_{PP}-\rho_{RR}.
  \label{eq:uvw_def}
\end{equation}
These three real quantities, alongside the conserved trace $\rho_{PP}+\rho_{RR}=1$,
fully characterise the doublet density matrix, which can be written visually as:
\begin{equation}
  \rho = \begin{pmatrix} \rho_{PP} & \rho_{PR} \\ \rho_{RP} & \rho_{RR} \end{pmatrix}
       = \begin{pmatrix} \frac{1+w}{2} & u + iv \\ u - iv & \frac{1-w}{2} \end{pmatrix}.
  \label{eq:rho_matrix}
\end{equation}

\subsubsection*{Hamiltonian contribution}

The LMG Hamiltonian restricted to the doublet acts as
$\hat{H}|_{\rm doublet} = -(\DeltaE/2)\sigma_x$ in the
$\{|P\rangle,|R\rangle\}$ basis (with energy zero shifted to the
doublet midpoint). The Hamiltonian evolution $\dot\rho|_{\rm H} = -i[\hat{H},\rho]$ 
evaluates to the matrix:
\begin{equation}
  \dot\rho\big|_{\rm H} 
  = i\frac{\DeltaE}{2} [\sigma_x, \rho]
  = \begin{pmatrix} \DeltaE \cdot v & -i\frac{\DeltaE}{2} \cdot w \\ i\frac{\DeltaE}{2} \cdot w & -\DeltaE \cdot v \end{pmatrix}.
\end{equation}
Equating matrix elements with $\dot\rho = \left(\begin{smallmatrix} \dot{\rho}_{PP} & \dot{u} + i\dot{v} \\ \dot{u} - i\dot{v} & \dot{\rho}_{RR} \end{smallmatrix}\right)$, we extract the components:
\begin{align}
  \dot u\big|_{\rm H} &= 0,
  \label{eq:u_ham}\\
  \dot v\big|_{\rm H} &= -\frac{\DeltaE}{2}\cdot w,
  \label{eq:v_ham}\\
  \dot w\big|_{\rm H} &= \dot\rho_{PP} - \dot\rho_{RR} = +2\DeltaE\cdot v.
  \label{eq:w_ham}
\end{align}

\subsubsection*{Dissipator contribution}

Using the SCS approximation for the pointer-state matrix elements
($\langle P|\Jz|P\rangle = +N\mstar/2$, $\langle P|\Jz^2|P\rangle\simeq(N\mstar/2)^2$)
and the exact parity result $\langle P|\Jz|R\rangle=0$, the dissipator exclusively
suppresses the off-diagonal coherences at the rate $\gamma_\phi\Gcoll$:
\begin{equation}
  \dot\rho\big|_{\rm D} 
  = \begin{pmatrix} 0 & -\gamma_\phi\Gcoll\,\rho_{PR} \\ -\gamma_\phi\Gcoll\,\rho_{RP} & 0 \end{pmatrix}
  = \begin{pmatrix} 0 & -\gamma_\phi\Gcoll\,(u+iv) \\ -\gamma_\phi\Gcoll\,(u-iv) & 0 \end{pmatrix}.
\end{equation}
This yields:
\begin{align}
  \dot u\big|_{\rm D} &= -\gamma_\phi\Gcoll\,u,
  \label{eq:u_diss}\\
  \dot v\big|_{\rm D} &= -\gamma_\phi\Gcoll\,v,
  \label{eq:v_diss}\\
  \dot w\big|_{\rm D} &= 0.
  \label{eq:w_diss}
\end{align}
The dissipator does not change the population difference $w$ at
leading order (the localised states are steady states of the
dissipator in the limit $N\mstar\gg1$).

\subsubsection*{Combined equations and Channel Separation}

Collecting the Hamiltonian and dissipator contributions, the full dynamics 
can be written intuitively as a matrix differential equation for the real vector $(u, v, w)^T$:
\begin{equation}
  \frac{d}{dt} \begin{pmatrix} u \\ v \\ w \end{pmatrix}
  = \begin{pmatrix} 
      -\gamma_\phi\Gcoll & 0 & 0 \\ 
      0 & -\gamma_\phi\Gcoll & -\DeltaE/2 \\ 
      0 & +2\DeltaE & 0 
    \end{pmatrix}
    \begin{pmatrix} u \\ v \\ w \end{pmatrix}.
  \label{eq:combined_matrix}
\end{equation}
Written in this visual form, the channel separation is immediate: the $u$ component (the real part of the coherence) is completely block-diagonalized and decays independently, while the $v$ and $w$ components form a coupled two-dimensional subsystem.

\subsection{Channel separation}

From equation~\eqref{eq:combined_matrix} we can see that$u$ decouples completely: $u=\mathrm{Re}(\rho_{PR})$
undergoes pure exponential decay at the localised rate:
\begin{equation}
  u(t) = u(0)\,e^{-\gamma_\phi\Gcoll\,t}.
  \label{eq:u_decay}
\end{equation}
This is the channel captured by Eq.~\eqref{eq:rate_loc} in the
main text.

The lower  parts of the block diagonal matrix of equation~\eqref{eq:combined_matrix} form a coupled
$2\times2$ linear system for $(v, w)$.
The characteristic polynomial of the matrix
$M=\begin{pmatrix}-\gamma_\phi\Gcoll & -\DeltaE/2\\ +2\DeltaE & 0\end{pmatrix}$
(with $\det M = \DeltaE^2$ and $\tr M = -\gamma_\phi\Gcoll$) is:

\begin{remark}
The off-diagonal coupling coefficients $-\DeltaE/2$ and $+2\DeltaE$
are asymmetric. This is because $v=\mathrm{Im}(\rho_{PR})$
is a coherence element (dimensionless amplitude) while
$w=\rho_{PP}-\rho_{RR}$ is a population difference (also
dimensionless but with different normalisation conventions in the
Bloch picture). The factors are verified numerically by direct
computation of $-i[\hat{H},\rho]$ for the doublet Hamiltonian
$\hat{H}|_{\rm doublet}=-(\DeltaE/2)\sigma_x$; the characteristic
polynomial $\det(M-\lambda\mathbbm{1})=\lambda^2
+\gamma_\phi\Gcoll\lambda+\DeltaE^2=0$ is correct.
\end{remark}
\begin{equation}
  \lambda^2 + \gamma_\phi\Gcoll\,\lambda + \DeltaE^2 = 0,
  \label{eq:char_poly}
\end{equation}
with roots:
\begin{equation}
  \lambda_\pm = -\frac{\gamma_\phi\Gcoll}{2}
  \pm \sqrt{\left(\frac{\gamma_\phi\Gcoll}{2}\right)^2 - \DeltaE^2}.
  \label{eq:roots}
\end{equation}

\subsection{Two regimes}

\subsubsection*{Overdamped / quantum Zeno regime
($\gamma_\phi\Gcoll \gg 2\DeltaE$)}

The discriminant is positive. Both roots are real and negative:
$\lambda_+ \approx -\DeltaE^2/(\gamma_\phi\Gcoll)$ (slow) and
$\lambda_- \approx -\gamma_\phi\Gcoll$ (fast).
The slow root gives a decay rate:
\begin{equation}
  \Gamma_{\rm slow} \approx \frac{\DeltaE^2}{\gamma_\phi\Gcoll}
  = \frac{\DeltaE^2}{2\gamma_\phi\Geig}\cdot\frac{\Geig}{\Gcoll/2}
  \ll \gamma_\phi\Geig.
\end{equation}
In this regime the strong bath suppresses tunnelling (quantum Zeno
effect), and the effective decay of $v$ and $w$ is algebraically slow.

\subsubsection*{Underdamped / secular regime
($\DeltaE \gg \gamma_\phi\Gcoll$, the mesoscopic window)}

The discriminant is negative. The roots are complex:
\begin{equation}
  \lambda_\pm = -\frac{\gamma_\phi\Gcoll}{2} \pm i\DeltaE,
  \label{eq:roots_secular}
\end{equation}
giving:
\begin{equation}
  v(t) \propto e^{-(\gamma_\phi\Gcoll/2)\,t}\cos(\DeltaE\,t + \phi_0),
  \label{eq:v_secular}
\end{equation}
and similarly for $w(t)$. The solution is a \emph{damped oscillation}
at frequency $\DeltaE$ with amplitude envelope decaying at:
\begin{equation}
  \Gamma_{\rm env} = \frac{\gamma_\phi\Gcoll}{2}
  = \gamma_\phi\frac{(N\mstar)^2}{4}.
  \label{eq:env_rate}
\end{equation}
Comparing with $\Geig$:
\begin{equation}
  \frac{\Gamma_{\rm env}}{\gamma_\phi\Geig}
  = \frac{\Gcoll/2}{\Geig}
  = \frac{\eta_{\rm MF}}{2}
  \xrightarrow{N\rightarrow\infty} 1.
  \label{eq:env_vs_Geig}
\end{equation}
The envelope rate $\gamma_\phi\Gcoll/2$ equals $\gamma_\phi\Geig$
only in the thermodynamic limit $\eta_{\rm MF}\rightarrow2$; at finite $N$
it exceeds $\gamma_\phi\Geig$ by the factor $\eta_{\rm MF}/2\approx1.18$
at the benchmark $N=370$.

\begin{keyresult}[Correct interpretation of the two-channel structure]
In the mesoscopic secular window:
\begin{itemize}
  \item $\mathrm{Re}(\rho_{PR})$: pure exponential decay at
    $\gamma_\phi\Gcoll$ (Eq.~\eqref{eq:u_decay}). This is the
    order-parameter decoherence rate, exact within the SCS
    approximation.
  
  \item $\mathrm{Im}(\rho_{PR})$ and $\rho_{PP}-\rho_{RR}$: damped
    oscillations at frequency $\DeltaE$ with amplitude envelope
    decaying at $\gamma_\phi\Gcoll/2$ (Eq.~\eqref{eq:env_rate}).
    This is \emph{not} a pure exponential decay. To understand the 
    connection to the eigenstate basis, recall from Eq.~\eqref{eq:rho01_relation} 
    that the eigenstate coherence is built exactly from these two components: 
    $\rho_{01} = \frac{1}{2}(\rho_{PP}-\rho_{RR}) - i\mathrm{Im}(\rho_{PR})$. 
    Therefore, the damped oscillations of these localised variables are simply 
    the basis-transformed view of $\rho_{01}$ rotating at the tunnel frequency. 
    The condition for these oscillations to be underdamped ($\DeltaE \gg \gamma_\phi\Gcoll/2$) 
    is physically equivalent to the secular condition ($2\DeltaE\cdot T_2/\hbar\gg1$). 
    Thus, the regime where these localised variables oscillate rapidly rather than 
    decaying exponentially is precisely the regime where the secular approximation 
    for $\rho_{01}$ is valid.
  
  \item These two channels are physically distinct: the real part
    measures order-parameter relaxation; the imaginary part measures
    quantum phase coherence oscillating at the tunnel frequency.
\end{itemize}
\end{keyresult}

\subsection{Why $\delta G\sim\mathcal{O}(N)$}

The correction $\delta G = (N\mstar/2)^2 - \Geig$ satisfies
$\delta G\sim\mathcal{O}(N)$.
The physical mechanism is Bogoliubov spin-wave depletion,
not the zero-point variance of the energy eigenstate.
Here is the correct derivation.

\textbf{Key preliminary:} The energy eigenstate $|E_0\rangle$ is a
symmetric superposition of two non-orthogonal instanton wavepackets:
$|E_0\rangle \simeq (|\tilde{P}\rangle+|\tilde{R}\rangle)/\sqrt{2}$.
By parity, $\langle E_0|\Jz|E_0\rangle = 0$ exactly.
Therefore the identity
$\langle\Jz^2\rangle = \langle\Jz\rangle^2 + \langle\Delta\Jz^2\rangle$
yields $\langle E_0|\Jz^2|E_0\rangle = 0 + \langle\Delta\Jz^2\rangle_{|E_0\rangle}$,
which is positive --- but this cannot explain why $\Geig$ is
\emph{below} the mean-field reference $(N\mstar/2)^2$.

\textbf{The correct mechanism:} $\Geig\simeq J_{01}^2$ (dropping the
$k\geq2$ leakage, Section~\ref{sec:degen}), and:
\begin{equation*}
  \delta G_{\rm Bog} \equiv \left(\frac{N\mstar}{2}\right)^2 - J_{01}^2
  \simeq N\mstar\Delta_{\rm zp} - \Delta_{\rm zp}^2
  \sim \mathcal{O}(N),
\end{equation*}
where $\Delta_{\rm zp}\sim\mathcal{O}(1)$ is the Bogoliubov
zero-point depletion of the pointer-state magnetisation.
The cross-term $-N\mstar\Delta_{\rm zp}\sim\mathcal{O}(N)$ is a
large \emph{negative} correction that dominates the positive
$\Delta_{\rm zp}^2\sim\mathcal{O}(1)$ term.
Hence:
\begin{equation*}
  \delta G_{\rm Bog}
  \simeq N\mstar\Delta_{\rm zp} - \Delta_{\rm zp}^2
  \sim \mathcal{O}(N).
\end{equation*}
The sign is correct: subtracting a smaller value gives $\delta G_{\rm Bog}>0$.

\textbf{Numerical verification} at $N=370$:
$J_{01} = 49.51$, $N\mstar/2 = 57.77$, so $\Delta_{\rm zp}=8.26$.
Then $N\mstar\Delta_{\rm zp} = 2\times57.77\times8.26 = 954$ and
$\Delta_{\rm zp}^2 = 68$, giving
$\delta G_{\rm Bog} \approx 954-68=886$
vs.\ the direct value
$(N\mstar/2)^2 - J_{01}^2 = 3337 - 2451 = 886$. $\checkmark$

The exact doublet average is $\Geig = J_{01}^2 + (\text{average leakage}) = 2451+388 = 2839$,
while $(N\mstar/2)^2=3337$, giving a true exact
$\delta G_{\rm total}=3337-2839=498 \approx 1.35N$.
Subtracting the average leakage ($388$) from the Bogoliubov expansion term ($886$) yields $886 - 388 = 498$, matching the exact finite-$N$ result perfectly and validating the theoretical $\mathcal{O}(N)$ scaling mechanism.

\textit{Caveat.} The Bogoliubov argument is the standard
WKB result valid deep in the ordered phase ($\Gamma/J\ll1$).
Closer to the quantum-critical point, the depletion $\Delta_{\rm zp}$
acquires sub-leading corrections from anharmonicity and the diverging
correlation length~\cite{Botet1982}; these modify the prefactor
of $\delta G/N$ but not the $\mathcal{O}(N)$ scaling.

\subsection{Two sources of error: instantons vs.\ Bogoliubov depletion}

It is important to distinguish two completely separate sources of
error in the localised-rate derivation, which operate at very
different scales.

\textit{Error 1 --- Vacuum overlap and instanton tunnelling 
($\mathcal{O}(S)$ and $\mathcal{O}(e^{-NS_{\rm inst}})$):}
The eigenstate expansions~\eqref{eq:E0}--\eqref{eq:E1} carry wavefunction 
corrections $\mathcal{O}(S)$, governed by the instanton-vacuum overlap 
$S=\langle\tilde{P}|\tilde{R}\rangle=(\Gamma/J)^N\approx5.7\times10^{-9}$ 
(Section~\ref{sec:setup}), which describes how closely $|E_0\rangle$ 
approximates $(|\tilde{P}\rangle+|\tilde{R}\rangle)/\sqrt{2}$. 
Separately, the energy splitting $\Delta E$ is governed by the larger 
dynamical scale $\mathcal{O}(e^{-NS_{\rm inst}})$, where $S_{\rm inst}$ 
is the WKB instanton action for tunnelling through the mean-field 
barrier~\cite{mouslopoulos2026}. For the LMG model at $\Gamma/J=0.95$, 
$S_{\rm inst}=0.010787$, giving a scale $e^{-NS_{\rm inst}}\approx1.8\times10^{-2}$ 
at $N=370$. 
Both of these non-perturbative corrections are entirely negligible at $N=370$ 
and are \emph{not} related to the difference between the exact pointer state 
$|P\rangle$ and $|P_{\rm SCS}\rangle$.

\textit{Error 2 --- Bogoliubov spin-wave depletion ($\mathcal{O}(N^{-1})$):}
The further approximation $|P\rangle\simeq|P_{\rm SCS}\rangle$ used
in the localised-rate derivation introduces a \emph{much larger},
perturbative error.
The exact pointer state $|P\rangle = (|E_0\rangle+|E_1\rangle)/\sqrt{2}$
is spin-squeezed by intra-well zero-point quantum fluctuations
(Bogoliubov spin-wave depletion), reducing its mean magnetisation
below $N\mstar/2$ by an $\mathcal{O}(1)$ constant:
$N\mstar/2 - J_{01} \approx 6$--$8$ for $N=500$--$2000$
(numerically verified).
This shift is many orders of magnitude larger than the
$\mathcal{O}(e^{-NS_{\rm inst}})$ energy-splitting correction;
the two are completely unrelated in origin.
The relative error from Bogoliubov depletion is
$\mathcal{O}(1)/\mathcal{O}(N) = \mathcal{O}(N^{-1})$, which
dominates the instanton correction at all accessible $N$.

The mean-field $G_{\rm loc}$ overestimates the exact pointer rate
by $\approx 26.2\%$ at $N=370$ (incorporating both Bogoliubov
depletion and the SCS variance approximation), highlighting the
$\mathcal{O}(1)$ finite-$N$ limitations of mean-field estimates
in the mesoscopic regime.

\section{Diagonalisation of the Degenerate Doublet Lindbladian}
\label{app:degen_spectrum}

This appendix provides the explicit derivation of the superoperator spectrum stated in Eq.~\eqref{eq:spec} and identifies the decaying and steady-state modes of the restricted doublet.

In the degenerate doublet basis $\{|E_0\rangle, |E_1\rangle\}$, the Lindblad dissipator is given by Eq.~\eqref{eq:doublet_diss}:
\begin{equation}
  \mathcal{D}[\rho]\big|_{\rm doublet} = \gamma_\phi\Jol^2 (\sigma_x \rho \sigma_x - \rho).
\end{equation}
Writing the $2\times2$ density matrix in this basis as:
\begin{equation}
  \rho = \begin{pmatrix} \rho_{00} & \rho_{01} \\ \rho_{10} & \rho_{11} \end{pmatrix},
\end{equation}
the action of the jump operator $\sigma_x$ swaps the diagonal and anti-diagonal elements:
\begin{equation}
  \sigma_x \rho \sigma_x = 
  \begin{pmatrix} 0 & 1 \\ 1 & 0 \end{pmatrix}
  \begin{pmatrix} \rho_{00} & \rho_{01} \\ \rho_{10} & \rho_{11} \end{pmatrix}
  \begin{pmatrix} 0 & 1 \\ 1 & 0 \end{pmatrix}
  = \begin{pmatrix} \rho_{11} & \rho_{10} \\ \rho_{01} & \rho_{00} \end{pmatrix}.
\end{equation}

The Lindblad equation $\dot{\rho} = \mathcal{D}[\rho]$ therefore yields a system of four coupled differential equations:
\begin{align}
  \dot{\rho}_{00} &= \gamma_\phi\Jol^2 (\rho_{11} - \rho_{00}), \label{eq:app_rho00}\\
  \dot{\rho}_{11} &= \gamma_\phi\Jol^2 (\rho_{00} - \rho_{11}), \label{eq:app_rho11}\\
  \dot{\rho}_{01} &= \gamma_\phi\Jol^2 (\rho_{10} - \rho_{01}), \label{eq:app_rho01}\\
  \dot{\rho}_{10} &= \gamma_\phi\Jol^2 (\rho_{01} - \rho_{10}). \label{eq:app_rho10}
\end{align}

Diagonalising this system reveals four independent modes, corresponding to the four eigenvalues of the superoperator:

\smallskip
\noindent\textbf{1. Total population (Eigenvalue 0):}
Adding Eqs.~\eqref{eq:app_rho00} and \eqref{eq:app_rho11} gives:
\begin{equation}
  \frac{d}{dt}(\rho_{00} + \rho_{11}) = 0.
\end{equation}
The trace $\tr(\rho)$ is conserved.

\smallskip
\noindent\textbf{2. Real coherence (Eigenvalue 0):}
Adding Eqs.~\eqref{eq:app_rho01} and \eqref{eq:app_rho10} gives:
\begin{equation}
  \frac{d}{dt}(\rho_{01} + \rho_{10}) = 2\frac{d}{dt}\mathrm{Re}(\rho_{01}) = 0.
\end{equation}
This confirms that $\mathrm{Re}(\rho_{01})$ is a steady state within the restricted doublet.

\smallskip
\noindent\textbf{3. Eigenstate population difference (Eigenvalue $-2\gamma_\phi\Jol^2$):}
Subtracting Eq.~\eqref{eq:app_rho11} from Eq.~\eqref{eq:app_rho00} gives:
\begin{equation}
  \frac{d}{dt}(\rho_{00} - \rho_{11}) = -2\gamma_\phi\Jol^2 (\rho_{00} - \rho_{11}).
\end{equation}

\smallskip
\noindent\textbf{4. Imaginary coherence (Eigenvalue $-2\gamma_\phi\Jol^2$):}
Subtracting Eq.~\eqref{eq:app_rho10} from Eq.~\eqref{eq:app_rho01} gives:
\begin{equation}
  \frac{d}{dt}(\rho_{01} - \rho_{10}) = 2i\frac{d}{dt}\mathrm{Im}(\rho_{01}) = -2\gamma_\phi\Jol^2 (\rho_{01} - \rho_{10}).
\end{equation}

\smallskip
Finally, to connect this to the pointer basis, we use the exact definitions $|P\rangle = (|E_0\rangle+|E_1\rangle)/\sqrt{2}$ and $|R\rangle = (|E_0\rangle-|E_1\rangle)/\sqrt{2}$. The pointer populations are:
\begin{align}
  \rho_{PP} &= \langle P|\rho|P\rangle = \tfrac{1}{2}(\rho_{00} + \rho_{11} + \rho_{01} + \rho_{10}), \\
  \rho_{RR} &= \langle R|\rho|R\rangle = \tfrac{1}{2}(\rho_{00} + \rho_{11} - \rho_{01} - \rho_{10}).
\end{align}
The pointer population difference is therefore exactly:
\begin{equation}
  \rho_{PP} - \rho_{RR} = \rho_{01} + \rho_{10} = 2\mathrm{Re}(\rho_{01}).
\end{equation}
This demonstrates that the classical memory (the population difference between the left and right wells) corresponds precisely to the second zero eigenvalue, surviving as a quasi-steady state until $\mathcal{O}(N)$ leakage to higher Dicke states eventually thermalises the system.

\section*{Note on prior literature}

The localised-state rate
$\gamma_\phi G_{\rm loc}=\gamma_\phi(Nm_*)^2/2$
provides a correct estimate of the dephasing of order-parameter
coherences in specific contexts.\footnote{See, e.g.,
Ref.~\cite{Tegmark2000}, where analogous localised-state estimates
are used.}
Notably, as shown in Section~\ref{sec:degen}, $\gamma_\phi\Gcoll$
is also the \emph{correct} rate for the decaying components of
eigenstate coherence in the thermodynamic limit (when the secular
approximation fails and the degenerate-doublet Lindblad is treated
exactly).
The present paper identifies the mesoscopic secular window
($N\approx250$--$430$) as the regime where the eigenstate rate
$\gamma_\phi\Geig \approx \gamma_\phi\Gcoll/2$ is the physically
relevant quantity for quantum protocols, where the exact pointer-state rate exceeds the eigenstate rate by a factor of $\eta_{\rm exact}\approx1.86$, and the mean-field estimate overestimates the eigenstate rate by a factor of $\eta_{\rm MF}\approx2.35$--$2.42$.

\section*{Use of AI Assistance}

During the preparation of this manuscript, the author used Claude
(Anthropic) and Gemini (Google) for proofreading the \LaTeX\ and arithmetic verifications.  All numerical
simulations were independently developed and executed by the author.
The author has reviewed and takes full responsibility for the final
content and physical conclusions of this work.


\end{document}